\documentclass[paper=a4,fontsize=11pt,DIV=18]{scrartcl}
\pdfoutput=1

\usepackage[utf8]{inputenc}
\usepackage[T1]{fontenc}
\usepackage{lmodern}

\usepackage[ngerman, american]{babel}							
\usepackage[autostyle]{csquotes}

\usepackage[footnotesize]{caption}[2008/08/24]
\usepackage{amsmath,amsfonts,amssymb,mathtools}

\usepackage{braket}
\usepackage[separate-uncertainty=true]{siunitx}
\usepackage{graphicx}
\usepackage{epstopdf}

\usepackage{color}
\usepackage[svgnames,dvipsnames]{xcolor}

\usepackage[numbers,sort&compress]{natbib}

\addto\captionsamerican{

}

\usepackage{authblk}

\newcommand{\boldnabla}{\mbox{\boldmath$\nabla$}} 
\newcommand{\vect}[1]{\boldsymbol{#1}} 
\newcommand{\dInt}[1]{{\, \mathrm{d} {#1}}}

\newcommand{\ia}{\text{a}}
\newcommand{\el}{\text{e}}
\newcommand{\ab}{{\alpha\beta}}
\newcommand{\const}{ \text{const.} }
\newcommand{\alphaLC}{{\tilde{\alpha}}}
\newcommand{\betaLC}{{\tilde{\beta}}}
\newcommand{\gammaLC}{ {\tilde{\gamma}}}

\newcommand{\new}[1]{\textcolor{black}{#1}}

\usepackage{tikz}
\usepackage{helvet,times}
\usepackage{bm,textcomp}

\usepackage{braket}

\usepackage{pdfpages}

\usepackage[colorlinks=true, pdfborder={0 0 0}, plainpages=false, citecolor=teal, linkcolor=black]{hyperref} 
\usepackage{doi}

\hypersetup{
	pdftitle={Computational modeling of active deformable membranes embedded in 3D flows},
	pdfauthor={C. B\"acher and S. Gekle},
	pdfkeywords={active membranes; active gel theory; elastic thin shells; Immersed-Boundary method; Lattice-Boltzmann method; cytoskeleton}
}

\begin{document}

\setcounter{page}{1}

\title{Computational modeling of active deformable membranes embedded in three-dimensional flows}
\date{}
\author[1]{Christian B\"acher \thanks{christian.baecher@uni-bayreuth.de}}
\author[1]{Stephan Gekle \thanks{stephan.gekle@uni-bayreuth.de}}

\affil[1]{Biofluid Simulation and Modeling, Theoretische Physik VI, Universität Bayreuth, Bayreuth, Germany}

\maketitle 

\section*{Abstract}

Active gel theory has recently been very successful in describing biologically active materials such as actin filaments or moving bacteria in temporally fixed and simple geometries such as cubes or spheres.
Here we develop a computational algorithm to compute the dynamic evolution of an arbitrarily shaped, deformable thin membrane of active material embedded in a 3D flowing liquid. 
For this, our algorithm combines active gel theory with the classical theory of thin elastic shells.
To compute the actual forces resulting from active stresses, we apply a parabolic fitting procedure to the triangulated membrane surface.
Active forces are then dynamically coupled via an Immersed-Boundary method to the surrounding fluid whose dynamics can be solved by any standard, e.g. Lattice-Boltzmann, flow solver.
We validate our algorithm using the Green's functions of [Berthoumieux et al. \textit{New J. Phys.} \textbf{16}, 065005 (2014)] for an active cylindrical membrane subjected (i) to a locally increased active stress and (ii) to a homogeneous active stress.
For the latter scenario, we predict in addition a so far unobserved non-axisymmetric instability.
We highlight the versatility of our method by analyzing the flow field inside an actively deforming cell embedded in external shear flow.
Further applications may be cytoplasmic streaming or active membranes in blood flows.
\textbf{Published as: \textit{Phys. Rev. E} 99(6), 062418 (2019)}

\section{Introduction}

Many biological processes such as cell division or locomotion depend on the ability of living cells to convert chemical energy into mechanical work \cite{AlbertsMolecularBiologyCell2007}.
A prominent mechanism to achieve such a conversion are motor proteins which perform work through a relative movement of cross-linked cytoskeletal filaments.
This movement induces active stresses in the cell cortex 
\cite{clark_stresses_2014,berthoumieux_active_2014,saha_determining_2016,hiraiwa_role_2016,salbreux_mechanics_2017-1,ravichandran_enhanced_2017} 
which are transmitted via anchor proteins to the plasma membrane separating the interior of the cell from its surroundings.
Active stresses have an inherent non-equilibrium character and are the basis of physically unique processes in active fluid layers such as instabilities \cite{liverpool_instabilities_2003,sarkar_role_2015}, the emergence of spontaneous flows \cite{ramaswamy_activity_2016} or the creation of geometrical structures \cite{Hannezo_2015,gowrishankar_nonequilibrium_2016}.
They are furthermore responsible for large shape deformations during cell morphogenesis \cite{GrillGrowingstressfulbiophysical2011,salbreux_actin_2012,callan-jones_cortical_2016}, cell division  \cite{salbreux_hydrodynamics_2009,sedzinski_polar_2011,green_cytokinesis_2012,mendes_pinto_force_2013,turlier_furrow_2014,sain_dynamic_2015}, cell locomotion 
\cite{carlsson_mechanisms_2011,shao_coupling_2012,Marthmechanismcellmotility2015,callan-jones_actin_2016,campbell_computational_2017},
cell rheology \cite{fischer-friedrich_rheology_2016,fischer-friedrich_active_2018}
or spike formation on artificial vesicles \cite{keber_topology_2014}.

In recent years a theoretical framework has been developed describing cytoskeleton and motor proteins together as an active continuous gel \cite{kruse_generic_2005,julicher_active_2007,ramaswamy_mechanics_2010,marchetti_hydrodynamics_2013,prost_active_2015,ahmed_active_2015,fodor_how_2016,needleman_active_2017,yeomans_natures_2017,julicher_hydrodynamic_2018}.
This active gel theory in general treats the cell cortex as a viscoelastic medium with additional active contributions.
On time scales short compared to the viscoelastic relaxation time active gel theory can be formulated in the elastic limit \cite{berthoumieux_active_2014}.
For thin active 2D membranes embedded in 3D space, active gel theory can be reformulated into force balance equations using the formalism of differential geometry \cite{berthoumieux_active_2014, salbreux_mechanics_2017-1}.
Any active stress in the membrane is then balanced by a counter-stress, usually due to viscous friction, from the external medium.
The force balance equations can be applied on fixed membrane geometries such as spheres, cylinders or flat layers.
This often results in the prediction of regions in parameter space where the prescribed shape is expected to become unstable \cite{salbreux_hydrodynamics_2009,hannezo_mechanical_2012,keber_topology_2014,Hannezo_2015,Zhang_2016_activeVesicle,Saw_2017}.
Despite being a powerful qualitative tool, such calculations cannot make statements about the precise shape of the active membrane after the instability has set in.

To obtain actual shape predictions, a number of works start instead from a parametrized, free membrane shape which is adjusted so as to fulfill the force balance equations for a given set of parameters and boundary conditions.
This procedure can be carried out either analytically \cite{maitra_activating_2014,berthoumieux_active_2014} or numerically \cite{turlier_furrow_2014,sain_dynamic_2015,RuprechtCorticalContractilityTriggers2015,callan-jones_cortical_2016,reymann_cortical_2016,whitfield_immersed_2016}.
For example, \citet{berthoumieux_active_2014} derived Green's functions to predict the deformation of an infinitely long cylindrical active, elastic membrane resulting from the application of a point active stress.
For certain parameter ranges, these Green's functions exhibit divergences which have been interpreted as mechanical (buckling or Rayleigh-Plateau-like) instabilities of the cylindrical membrane.
\citet{sain_dynamic_2015} investigated the axisymmetric dynamics of the furrow in cytokinesis. 
\citet{turlier_furrow_2014} computed the time evolution of an axisymmetric membrane undergoing cytokinesis by advecting tracer points discretizing the membrane.
\citet{callan-jones_cortical_2016} predicted a transition to a polarized cell shape because of an instability of the cell cortex.
\citet{reymann_cortical_2016} matched axisymmetric, theoretical results to observed cell shapes during cytokinesis using measured velocity and order parameter fields as input for the theory. 
\citet{heer_actomyosin-based_2017} determined the equilibrium shape of an elastic tissue layer folded into a deformable ellipsoidal shell, where myosin activity is incorporated as a preferred curvature of the shell.
\citet{klughammer_cytoplasmic_2018} analytically calculated the flow inside a sphere that slightly deforms due to a traveling band of surface tension, which mimics cortical active tension, under the assumption of rotational symmetry.
In all of these works the full dynamics of the external fluid was neglected and nearly all of them restricted themselves to deformations from simple rest shapes in the steady state.
There currently exists no analytical or numerical method to compute the dynamical deformation of an arbitrarily shaped active membrane immersed in a 3D moving Newtonian fluid.

In this work, we develop a computational algorithm to predict the dynamic deformation of arbitrarily shaped thin membranes discretized by a set of nodes connected via flat triangles.
Starting from prescribed active stresses on the discretized membrane our algorithm computes the corresponding forces on every membrane node via a parabolic fitting procedure taking into account the full deformed surface geometry.
Knowledge of the nodal forces then enables the dynamic coupling to a surrounding 3D fluid via the Immersed-Boundary Method (IBM).
The Navier-Stokes dynamics for the surrounding fluid is solved here by the Lattice-Boltzmann method (LBM), but other flow solvers can straightforwardly be incorporated.
With this, our method allows to study the dynamic evolution of active membranes in general external flows. 
It thus builds a bridge between the extensive literature on active fluids and the similarly extensive work on elastic cells, vesicles and capsules in flows \cite{barthes-biesel_modeling_2011,freund_numerical_2014,farutin_3d_2014,guckenberger_bending_2016,secomb_blood_2017,balogh_computational_2017,bacher_clustering_2017, guckenberger_numericalexperimental_2018}.
In biological situations often a flowing environment is present rendering the dynamic coupling of external fluid and membrane deformations necessary.
Our proposed method allows such a coupling and thus the computation of dynamically evolving non-equilibrium shapes.
Possible applications of our method include the study of active cell membranes inside the blood stream or active cellular compartments in cytoplasmic streaming flows.

First in section \ref{SEC:couplingMemFluid}, we extensively describe the LBM and IBM for a dynamic coupling of an elastic membrane and a suspending fluid.
In section \ref{SEC:forceBalance} we start with the problem formulation in the framework of thin shell theory using differential geometry.
Next, we describe our algorithm for three dimensional active force calculation in section \ref{SEC:algorithm}.
In section~\ref{SEC:validation} we provide an in-depth validation of our algorithm based on the analytical results for a cylindrical active membrane in case of infinitesimal deformations by \citeauthor{berthoumieux_active_2014} \cite{berthoumieux_active_2014}.
As an application, section \ref{SEC:FlowInSpheroid} presents the flow field inside a dividing elastic cell, where active stresses trigger membrane deformations that in turn lead to fluid flow.
In section \ref{SEC:DividingShearFlow} we analyze the same system in externally driven shear flow.
Eventually, we conclude our work in section \ref{SEC:conclusion}.

\section{Dynamic coupling of membrane and fluid}
\label{SEC:couplingMemFluid}

\subsection{Lattice-Boltzmann method for fluid dynamics}
\label{SEC:LBM}

The Lattice-Boltzmann method (LBM) is an efficient and accurate flow solver which is well described in the literature \cite{succi_lattice_2001, DunwegLatticeBoltzmannsimulations2009, aidun_lattice-boltzmann_2010,kruger_lattice_2016}.
In the following we therefore summarize only the basic concepts.
In contrast to macroscopic, e.g. finite element, methods based directly on the discretized Navier-Stokes equation (NSE), LBM starts from the Boltzmann equation which is a common tool in statistical mechanics.
Using Chapman-Enskog analysis the NSE are recovered from the Boltzmann equation \cite{harris_introduction_2004,kruger_lattice_2016}.

The Boltzmann equation provides insight into a system on mesoscopic length scales by means of the continuous particle distribution function $f(\vect{r},\vect{p},t)$ where
$\vect{r}$, $\vect{p}$ and $t$ refer to position of the particles, momentum of the particles and time, respectively.
The expression $f(\vect{r},\vect{p},t) \dInt{\vect{r}} \dInt{\vect{p}} \dInt{t}$ gives the probability to find a particle (fluid molecule) in the phase-space volume $\dInt{\vect{r}} \dInt{\vect{p}}$ at $(\vect{r},\vect{p})$ in the time interval $t$ to $t+dt$.
The dynamic evolution of the particle distribution function $f(\vect{r},\vect{p},t)$ is given by
\begin{equation}
\frac{\dInt{f}}{\dInt{t}} = \frac{\partial f}{\partial t} + \vect{v} \cdot \boldnabla_{\vect{r}} f + \vect{F} \cdot \boldnabla_{\vect{p}} f = \Omega
\label{EQ:Boltzmann}
\end{equation}
with $\Omega$ being the collision operator accounting for re-distribution of molecules due to collisions.

\begin{figure}[h]
	a)	
	\begin{minipage}{0.45\textwidth}
		\centering
		\includegraphics[width=.7\textwidth]{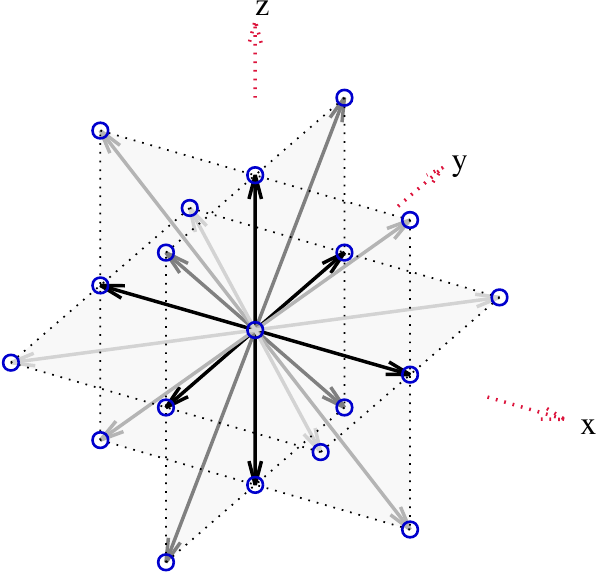}
	\end{minipage}
	b)
	\begin{minipage}{0.45\textwidth}
		\centering
		\includegraphics[width=\textwidth]{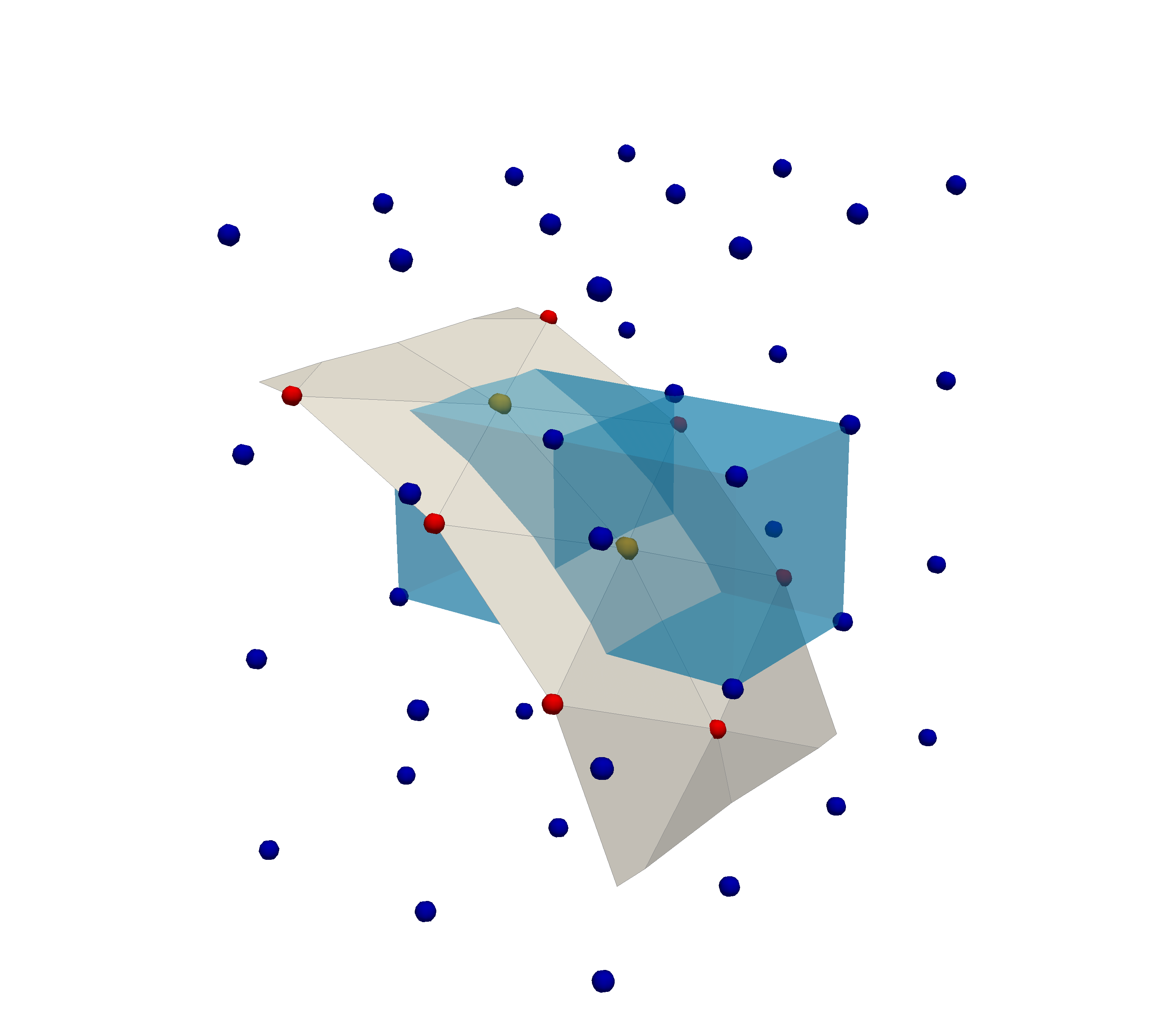}
	\end{minipage}
	
	\caption{
	a) D3Q19 LBM scheme with the discrete velocity set (solid arrows) connecting nearest and next nearest neighboring fluid nodes (dots).
	b) 3D illustration of the Immersed Boundary Method. 
	A continuous membrane is discretized by Lagrangian nodes (red and orange dots) that are connected by triangles.
	The membrane is immersed in an Eulerian grid representing the fluid (blue dots).
	The velocity at a Lagrangian node is obtained by interpolation from the eight closest fluid nodes (illustrated for the two orange membrane nodes in the middle by the blue shaded cubes).
	The same stencil is used to transmit the forces from the membrane to the fluid.
	}
	\label{FIG:IBMinterpolation}
\end{figure}

Discretization of space, momentum space and time leads from the Boltzmann equation \eqref{EQ:Boltzmann} to the Lattice-Boltzmann equation.
The discretization of the spatial domain is carried out using a cubic Eulerian grid. 
The distance between the fluid nodes is $\Delta x = 1$ in simulation units.
In contrast to other methods, LBM also discretizes the momentum (velocity) space,  i.e., $f = f(\vect{x}_j, \vect{p}_i,t)$ such that only a discrete set of velocities is allowed at each node.
Here, we use the common D3Q19 scheme with 19 discrete velocity vectors, which is illustrated in figure \ref{FIG:IBMinterpolation}~a).
Thus, each node contains one population $f(\vect{x}_j, \vect{p}_i,t)$ for each momentum $\vect{p}_i$, which moves with the corresponding velocity $\vect{c}_i$ away from the node $\vect{x}_j$ within one time step.
As an abbreviation the distribution functions are labeled by an index according to their discrete momentum/velocity, i.e., $f(\vect{x}_j, \vect{p}_i,t)  = f_i(\vect{x}_j,t)$.
Finally, discrete time steps from $t$ to $t + \Delta t$ are considered.
Under discretization the Boltzmann equation in \eqref{EQ:Boltzmann} becomes the Lattice-Boltzmann equation \cite{kruger_lattice_2016}
\begin{equation}
f_i(\vect{x}_j+\vect{c}_i \Delta t, t+\Delta t) = f_i(\vect{x}_j, t) + \Omega_i(\vect{x}_j, t) \Delta t.
\label{EQ:LatticeBoltzmann}
\end{equation}
The numerical integration of the Lattice-Boltzmann equation in time is split into two steps, collision and propagation.
Collision is done by an approximation of the collision operator in equation \eqref{EQ:LatticeBoltzmann}.
The idea for the approximation of the collision operator is that the populations should relax towards the Maxwellian equilibrium distribution function in absence of driving forces.
Here, we use the multiple relaxation time scheme.
In the framework of the multiple relaxation time scheme the relaxation is done in moment space of the populations with individual relaxation rates for the different moments \cite{DunwegLatticeBoltzmannsimulations2009,kruger_lattice_2016}.
\new{
Moments corresponding to conserved density and momentum do not have a relaxation time, while two relaxation rates related to bulk and shear viscosity, respectively, are chosen for moments corresponding to the stress tensor.
For further discussion on the relaxation rates and moments used in our LBM implementation we refer to refs. \cite{dunweg_statistical_2007,DunwegLatticeBoltzmannsimulations2009}.
}
In the second step, the streaming, the populations after collision propagate according to the associated velocities.

\new{
We note that our LBM implementation supports adding thermal fluctuations to the fluid dynamics mimicking a given thermal energy $k_B T$ 
 corresponding to fluctuating hydrodynamics \cite{dunweg_statistical_2007}.
Thermal fluctuations are taken into account by adding a random noise to those relaxed moments of the multiple relaxation time scheme which correspond to elements of the stress tensor \cite{dunweg_statistical_2007,DunwegLatticeBoltzmannsimulations2009}. 
We note that our approach using thermal fluctuations is different from considering a separate temperature field \cite{he_novel_1998,coelho_lattice_2014,coelho_lattice_2018} as e.g. required for convection flows \cite{he_novel_1998}.
Although they are not per se necessary, we employ thermal fluctuations in the present manuscript to speed up the onset of instabilities. 
}

\new{
Although we do not consider solid walls in the present manuscript, we note that they can be realized by bounce back boundary conditions \cite{kruger_lattice_2016}, where populations streaming towards nodes beyond a boundary are simply reflected.
In case of moving elastic objects (\ref{SEC:IBM}) having close contact to solid walls as it is considered e.g. in ref. \cite{zhang_sharpedged_2019}, at least one fluid node between the object and the solid boundary is required in our method.
}

Typical LBM grids used in this work have dimensions of 126x72x72 or 100x100x100.
Except in the case of the shear flow with moving boundaries in section \ref{SEC:DividingShearFlow}, our LBM simulations use periodic boundary conditions for the fluid.
A typical simulation run, e.g. in figure~\ref{FIG:bellShaped} consists of $10^{7}$ time steps.
Onset of Rayleigh-Plateau like instabilities in figure~\ref{FIG:PD_3D} typically appears after approximately $8\times10^{6}$ time steps.
Simulations are performed with the simulation package ESPResSo \cite{limbach_espressoextensible_2006,roehm_lattice_2012,arnold_espresso_2013} which has been extended to include thin membranes using the Immersed Boundary Method described next.

\subsection{Immersed-Boundary method for membrane dynamics}
\label{SEC:IBM}

The framework of the Immersed-Boundary Method (IBM) \cite{peskin_immersed_2002-1,mittal_immersed_2005-1,mountrakis_revisiting_2017} allows for the coupling of a cellular membrane to the suspending fluid, where the fluid is simulated using LBM as described in the previous section \ref{SEC:LBM}.
The IBM consists of two central steps: elastic forces acting on the membrane are transmitted to the fluid and due to no-slip boundary condition the mass-less membrane is advected with the local fluid velocity.

The membrane of a cell is represented by an infinitely thin elastic sheet in the framework of thin shell theory \cite{green_theoretical_1954,green_large_1960,koiter_mathematical_1970}.
In our numerical simulations the membranes is discretized by a set of nodes that are connected by flat triangles \cite{pozrikidis_interfacial_2001, farutin_3d_2014,seol_immersed_2016,guckenberger_theory_2017}.
These represent the elastic membrane as a Lagrangian mesh immersed into the Eulerian fluid mesh as illustrated in figure \ref{FIG:IBMinterpolation}~b).

Physical behavior of the membrane is characterized by appropriate constitutive laws which are described in section \ref{SEC:forceBalance}.
The resulting forces are calculated on the deformed Lagrangian membrane mesh as described in section~\ref{SEC:algorithm}.
To transmit these forces into the fluid, the incompressible NSE for the fluid velocity $\vect{u}(\vect{x},t)$ becomes \cite{seol_immersed_2016}
\begin{equation}
	\frac{\partial \vect{u}}{\partial t} + \left( \vect{u}\cdot\nabla \right) \vect{u} = -\frac{1}{\rho} \nabla p + \nu \Delta \vect{u} + \frac{1}{\rho}\int \vect{f}\left( \vect{X}^\prime,t \right) \delta\left( \vect{X}^\prime - \vect{x} \right) \dInt{^2 X^\prime},
\end{equation}
with $p$ the pressure, $\nu$ the kinematic viscosity, $\rho$ the density, and $\vect{f}$ the membrane force per area acting on the membrane, which is parametrized by $\vect{X}^\prime$.
The Dirac $\delta$ function ensures that the forces only act at the position of the actual membrane.
Correspondingly, the Lattice-Boltzmann equation \eqref{EQ:LatticeBoltzmann} changes to
\begin{align}
	\vect{F}_j &= \int \vect{f}\left( \vect{X}^\prime,t \right) \delta \left( \vect{X}^\prime - \vect{x}_j \right) \dInt{^2 X^\prime} \label{EQ:ForceLBM} \\
	f_i(\vect{x}_j+\vect{c}_i \Delta t &, t+\Delta t) = f_i(\vect{x}_j, t) + \Omega_i(\vect{x}_j, \vect{F}_j ,t) \Delta t. \label{EQ:LBMwithForceIBM}
\end{align} 

After an update of the fluid dynamics (LBM algorithm), which now include the forces from the membrane $\vect{F}_j$, the mass-less membrane nodes are advected with the local fluid velocity thus satisfying exactly the no-slip boundary condition \cite{mittal_immersed_2005-1}.
Moving with the local fluid velocity is expressed for a membrane node $\vect{x}_n \in \vect{X}^\prime$ by \cite{peskin_immersed_2002-1,mittal_immersed_2005-1}
\begin{equation}
	\frac{\dInt{\vect{x}_n}}{\dInt{t}} = \vect{u} \left( \vect{x}_n(t), t \right) = \int \vect{u}(\vect{x},t) \delta\left( \vect{x}-\vect{x}_n \right) \dInt{^3x}.
\label{EQ:IBM}
\end{equation}
with $\vect{u} \left( \vect{x}_n, t \right)$ being the fluid velocity at the position of the membrane node.
In simulations equation (\refeq{EQ:IBM}) is integrated numerically using an Euler scheme in order to move each membrane node from time $t$ to $t+\Delta t$.

The core of both steps, force transmission and movement with local fluid velocity, is the interpolation between the Eulerian fluid grid and the Lagrangian membrane grid.
Considering the discretization and the resulting spatial mismatch of membrane nodes and fluid nodes, as illustrated in figure \ref{FIG:IBMinterpolation}~b), it becomes clear that an interpolation is necessary.
On the one hand the ideal point force at the site of a membrane node must be spread to the adjacent fluid nodes.
On the other hand to obtain the local fluid velocity at the site of a membrane node the velocity of the adjacent fluid nodes is interpolated.
This means the $\delta$ distributions in equation \eqref{EQ:ForceLBM} and \eqref{EQ:IBM} must be discretized.
The interpolation between fluid and membrane is carried out by an eight-point stencil.
As illustrated for two membrane nodes by the blue shaded cubes in figure \ref{FIG:IBMinterpolation}~b),  for each membrane node a cube containing the eight nearest fluid nodes is considered.
A linear interpolation between the eight fluid points is performed \cite{ahlrichs_simulation_1999-1}.

A typical mesh for a cylinder in the present study contains 17100 Lagrangian membrane nodes and 34200 triangles.
Simulating a cylindrical membrane, as it is partly shown in figure \ref{FIG:delta3D}, is done using periodic boundary conditions for the fluid in all directions and for the membrane in the direction of the cylinder axis.
The latter is achieved by connecting membrane nodes at the end of the box to those at the beginning of the box.

\subsection{Validation of LBM/IBM for passive, elastic membranes}
\label{SEC:ValidationPassiveElastic}

Our implementation of the force calculation and the LBM/IBM have extensively been tested for passive, elastic membranes in previous publications \cite{guckenberger_bending_2016,guckenberger_theory_2017,quint_3d_2017,guckenberger_numericalexperimental_2018,gekle_strongly_2016,bacher_antimargination_2018}.

The algorithm for the passive, elastic force calculation has been validated in references \cite{guckenberger_bending_2016,guckenberger_theory_2017} by comparison with exact results in static situations as well as for a capsule in shear flow.
For red blood cells flowing through a rectangular channel very good agreement with \textit{in vitro} experiments has been found \cite{quint_3d_2017,guckenberger_numericalexperimental_2018}.

The mixed LBM/IBM has been extensively validated for suspensions of red blood cells and rigid particles in complex geometries.
In \cite{gekle_strongly_2016} the concentration profile of cells across the channel diameter (affected by cross streamline migration, which is triggered by the passive elasticity of the cells) has been successfully compared to well-established literature results.
In ref. \cite{bacher_antimargination_2018} we have performed a validation based on the Zweifach-Fung effect for red blood cell suspension in branching channels.
Furthermore, in the Supplementary Information of ref. \cite{bacher_antimargination_2018}  we have shown red blood cell shapes for a single cell in tube flow and in a rectangular channel, respectively, that are in very good agreement with previous studies using dissipative particle dynamics \cite{fedosov_deformation_2014} and boundary integral method \cite{guckenberger_numericalexperimental_2018}, respectively.
The stability and behavior of stiff particles realized by IBM has been shown and validated in ref. \cite{gekle_strongly_2016} and \cite{bacher_antimargination_2018} for a spherical particle based on the Stokes relation (sphere pulled through a quiescent fluid as well as a fixed sphere in homogeneous flow) and an ellipsoidal particle rotating in shear flow.

In addition to the well-established passive elasticity, in this paper we introduce active elastic forces into simulations of membranes, see section \ref{SEC:algorithm}.
We refer to section \ref{SEC:validation} for the validation in case of an active, elastic cell membrane.

\section{ Force balance in thin shell formulation including active stresses }

\label{SEC:forceBalance}

\begin{figure}
	\centering
	\includegraphics[width=\textwidth]{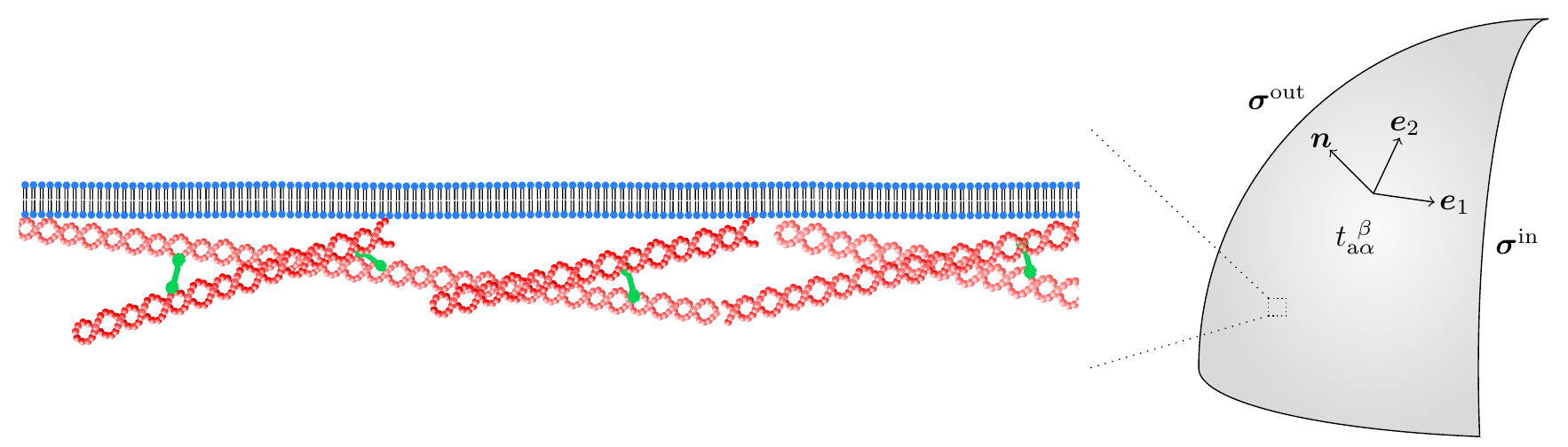} 
	\caption{ (left) The cell cortex underlying the plasma membrane consists of cytoskeletal filaments and motor proteins.
		The latter constantly convert energy into mechanical work.
		(right) Membrane and cortex are condensed into a thin shell with normal vector $\vect{n}$ and in-plane coordinates $\vect{e}_1$, $\vect{e}_2$.
		Mechanical work of motor proteins results in an active in-plane surface stress $t_{\ia\alpha}^{~\beta}$.
		Forces from the cytoplasm inside the cell and from the extra-cellular  medium onto the membrane can be treated by the 3D fluid stress tensors $\sigma_{ij}^{in}$ and $\sigma_{ij}^{out}$.	}
	\label{FIG:membraneSketch}
\end{figure}

\subsection{Physical model}

We consider the plasma membrane and the cytoskeletal filaments within the cell cortex, sketched in figure \ref{FIG:membraneSketch}, as a single physical entity which for simplicity we denote in the following as ''the membrane''.
Since this membrane is small in height compared to the cell diameter it is described as an infinitely thin shell.
Thin shell theory treats the membrane as a two dimensional manifold embedded in the three dimensional environment, e.g., the intra- and extra-cellular fluid, thereby accounting for membrane curvature.
For a condensed introduction into the required differential geometry on such manifolds as well as present conventions we refer the reader to ref.~\cite{deserno_fluid_2015}  and appendix~\ref{app:thinShell} of this work, respectively.
Here, we only note that Greek indices refer to coordinates on the membrane, i.e., $\alpha, \beta = 1,2$, and Einstein summation convention is used.
In this work, we consider the short-time limit of a purely elastic membrane noting that our algorithm can, without substantial difficulty, be extended to viscous or viscoelastic membranes.

The key quantity in the framework of thin shell theory are the in-plane surface stresses $t_\ab$ and moments $m_\ab$ \cite{green_theoretical_1954, green_large_1960, vetter_subdivision_2013}.
The in-plane surface stress tensor or stress resultant \cite{green_large_1960} $t_\ab$ is the 3D elastic stress tensor for the material forming the membrane
projected onto the membrane and integrated over the membrane thickness $h$ \cite{green_large_1960} (dimensions of a force per length, i.e., N/m)
\footnote{The term in-plane surface stresses mainly follows the terminology of \citet{deserno_fluid_2015}. \citet{guven_membrane_2004} and \citet{deserno_fluid_2015} denote the corresponding force vector with its in-plane components being $t_{\alpha\beta}$ as surface stress. For the same quantity \citet{berthoumieux_active_2014} and \citet{salbreux_mechanics_2017-1} as well as \citet{guckenberger_theory_2017} use the term in-plane tension tensor. \citet{daddi-moussa-ider_long-lived_2016} call this quantity stress tensor.  Finally, \citet{green_large_1960} call it stress resultant.}.
It is a function of in-plane coordinates $\alpha$, $\beta$ and contains only in-plane components such that its matrix representation has dimensions 2x2. 
Besides the in-plane surface stresses, we introduce the normal surface stress which is sometimes denoted a shearing force \cite{green_theoretical_1954} or transverse shear surface stress $t_n^\alpha$ \cite{pozrikidis_interfacial_2001,guckenberger_theory_2017}.
The in-place surface stresses $t_\ab$ contain a contribution from passive, elastic stresses as well as from active, force-generating mechanisms.
Both contributions superpose linearly and thus can be treated separately \cite{berthoumieux_active_2014,salbreux_mechanics_2017-1}.
Passive elastic stresses can be further split into different contributions such as, e.g., shear and bending resistances \cite{freund_numerical_2014,barthes-biesel_motion_2016, guckenberger_theory_2017}.
The moment tensor or stress couples \cite{green_large_1960} $m_\ab$ accounts for stress distribution across the membrane \cite{green_large_1960} (dimensions of a torque per unit length, i.e., N).
We denote the corresponding normal surface moment by $m_n$.
For explicit materials building up the membrane corresponding constitutive laws are required to derive explicit forms for in-plane surface stresses and moments.

Considering the passive, elastic properties of membranes in most cases constitutive laws are formulated in terms of a strain energy functional.
For many biological membranes such as red blood cells the passive, elastic forces arise from the resistance to shear deformation of the cortical, cytoskeletal network and the resistance to bending deformation and area dilatation of the plasma membrane.
For shear resistance and area dilatation  \citet{skalak_strain_1973} proposed an appropriate energy density functional
\begin{equation}
	w^{\text{SK}}(I_1, I_2) = \frac{\kappa_{\text{S}}}{12} \left( ( I_1^2 + 2I_1 - 2I_2)  + C I_2^2 \right).
	\label{EQ:Skalak}
\end{equation}
The energy density depends on the invariants $I_1, I_2$ of the transformation between undeformed and deformed membrane \cite{skalak_strain_1973, freund_numerical_2014}.
Both invariants are defined in equation \eqref{EQ:invariant1} and \eqref{EQ:invariant2} of the appendix.
Resistance to shear is characterized by the shear modulus $\kappa_{\text{S}}$ while resistance to area dilatation is characterized by $C \kappa_{\text{S}}$ with $C$ much larger than unity for nearly area-incompressible membranes.
For bending resistance the Helfrich model is used \cite{helfrich_elastic_1973,guckenberger_theory_2017}
\begin{equation} 
	w^{\text{HF}} = 2 \kappa_{\text{B}} (H-H_0)^2 + \kappa_K K
	\label{EQ:Helfrich}
\end{equation}
with $\kappa_{\text{B}}$ being the bending modulus, $H$ denoting the local mean curvature, $H_0$ the spontaneous curvature, $\kappa_K$ the Gaussian modulus and $K$ the Gaussian curvature.
On the one hand, from a given energy functional, in-plane surface stresses and moments can be deduced by functional derivatives with respect to the strain tensor \cite{green_large_1960,daddi-moussa-ider_long-lived_2016} and curvature tensor, respectively \cite{steigmann_fluid_1999,sauer_stabilized_2016,guckenberger_theory_2017}.
On the other hand, in numerical algorithms the explicit introduction of stresses is typically bypassed and forces on the nodes discretizing the membrane are often computed directly by deriving a discretized version of the energy functional, here equations \eqref{EQ:Skalak} and \eqref{EQ:Helfrich}, with respect to the node positions \cite{kraus_fluid_1996, freund_numerical_2014, guckenberger_theory_2017}.
This approach is used here as well for the passive elastic forces. 
For bending force calculation we use the method denoted by B in ref. \cite{guckenberger_bending_2016}.

For actively generated forces, however, this approach is not applicable since, due to their inherent non-equilibrium character, an energy functional cannot be defined in a strict sense.
Instead, active contributions are usually given in terms of active stresses whose strength and direction can either be temporally constant or depend on additional quantities such a local concentration of motor proteins \cite{salbreux_mechanics_2017-1}.
We will in the following construct a numerical method which explicitly applies these active stresses and derives the corresponding active forces on the discretized membrane nodes. 
These forces are then used to introduce a two-way coupling between active membrane dynamics and a surrounding hydrodynamic flow.
For simplicity of demonstration and in order to connect to the analytical axisymmetric solutions of \cite{berthoumieux_active_2014}, we restrict ourselves to temporally constant active in-plane surface stresses. 
Time-dependent active stresses or active moments can be included without substantial modification of our numerical framework.

\subsection{Force balance for deformable active membranes embedded in a 3D fluid}

As sketched in figure \ref{FIG:membraneSketch} we consider a membrane immersed in an external fluid and enclosing an internal fluid, the cytosol.
Coupling of the membrane to the internal and external fluid with the stress tensor $\sigma_{ij}^{in}$ and  $\sigma_{ij}^{out}$, respectively, is described by the traction jump \cite{pozrikidis_interfacial_2001,farutin_3d_2014,barthes-biesel_motion_2016}
\begin{equation}
 - f_j = \left( \sigma_{ij}^{out} - \sigma_{ij}^{in} \right) n_i
\end{equation}
with Latin subscripts denoting 3D coordinates and the components $n_i$ of the local unit normal vector onto the membrane pointing towards the external fluid.
Transforming to the in-plane coordinate system the traction jump can \-- as a vector on the membrane in general \-- be decomposed into tangential and normal component
\begin{equation}
  - \vect{f} = - f^\alpha \vect{e}_\alpha - f^n \vect{n}.
\end{equation}
Neglecting membrane inertia the traction jump between internal and external fluid is balanced by membrane forces (per unit area).
In the present situation, these arise from elastic and active contributions.
Looking ahead, the numerical method which we will construct in the following section will compute elastic forces $f_\el^\alpha$ and $f^n_\el$ via the classical discretized energy functional route, bypassing for simplicity the introduction of elastic stresses, while active forces need to be computed explicitly from active stresses and moments.
It is thus convenient to write the force balance for a thin shell \cite{green_theoretical_1954,berthoumieux_active_2014,daddi-moussa-ider_long-lived_2016,barthes-biesel_motion_2016,salbreux_mechanics_2017-1} as
\begin{align}
\nabla_{\alpha}^{\prime} t_\ia^{\alpha\beta} + C_{\alpha}^{\prime \beta} t_{n\ia}^{\alpha} + f_\el^{\beta}&= f^{\beta} \label{EQ:forceBalanceT} \\
\nabla_{\alpha}^{\prime} t_{n\ia}^{\alpha} - C^{\prime}_{\alpha\beta} t_\ia^{\alpha\beta} + f_\el^n&= f^n \label{EQ:forceBalanceN} \\
\nabla_{\alpha}^{\prime} m_\ia^{\alpha\beta} + C_{\alpha}^{\prime \beta} m_{n\ia}^{\alpha} &= \epsilon_{\alpha}^{\prime\beta} t_{n\ia}^{\alpha} \label{EQ:forceBalanceMT}\\
\nabla_{\alpha}^{\prime} m_{n\ia}^{\alpha}  - C^{\prime}_{\alpha\beta} m_\ia^{\alpha\beta} &= -\epsilon^\prime_{\alpha\beta}t_\ia^{\alpha\beta} \label{EQ:forceBalanceMN},
\end{align}
with active contributions to the in-plane and normal surface stress $t_\ia^{\alpha\beta},t_{n\ia}^{\alpha}$ and active contributions to the moments $m_\ia^{\alpha\beta},m_{n\ia}^{\alpha}$.
The geometrical quantities are the curvature tensor $C_{\alpha\beta}^{\prime}$, the Levi-Civita tensor $\epsilon^\prime_{\alpha\beta}$ and the co-variant derivative $\nabla^\prime_\alpha$ taken on the deformed membrane as defined in the appendix. 
On the basis of equation \eqref{EQ:forceBalanceT} and \eqref{EQ:forceBalanceN} it becomes clear that the negative traction jump is the force exerted from the membrane on the fluid.
We now consider the active contributions to the traction jump in the case of vanishing active moments ($m_\ia^{\alpha\beta}=m_{n\ia}^{\alpha}=0$)
and vanishing active transverse shear surface stress ($t_{n\ia}^{\alpha}=0$)
\begin{align}
f_\ia^{\beta}  &= \nabla_{\alpha}^{\prime} t_\ia^{\alpha\beta} = t^{\alpha\beta}_{\ia~,\alpha} + \Gamma_{\alpha\gamma}^{\prime\alpha} t_\ia^{\gamma\beta} + \Gamma_{\alpha\gamma}^{\prime\beta} t_\ia^{\alpha\gamma} \label{EQ:activeForceTang} \\
f_\ia^n &= - C^{\prime}_{\alpha\beta}t_\ia^{\alpha\beta},
\label{EQ:activeForceNorm}
\end{align}
where we have used the definition of the co-variant derivative on the membrane (see equation \eqref{EQ:covariantDeriv}) in the first line.
To simulate the temporal dynamics and coupling to the external fluid of our discretized active membrane, we need to compute the forces on each membrane node corresponding to active in-plane surface stresses.
According to equations~(\ref{EQ:activeForceTang}) and (\ref{EQ:activeForceNorm}) the curvature tensor, the Christoffel symbols, and the active in-plane surface stresses together with their derivatives have to be known locally on each node on a deformed surface.
In the next section we will develop an algorithm to compute these quantities numerically for the discretized thin shell embedded in a 3D environment.

\section{Algorithm for active force calculation on arbitrarily shaped discretized membranes}
\label{SEC:algorithm}

Our algorithm starts from active in-plane surface stresses and computes the corresponding active forces on the discretized membrane.
The key ingredient of the algorithm is the discrete computation of geometrical properties on the discretized, deformed membrane, such as the curvature tensor or Christoffel symbols.
This is achieved by a parabolic fitting procedure and using the force balance equations  \eqref{EQ:activeForceTang} and \eqref{EQ:activeForceNorm} as described in subsection \ref{SEC:ParabolicFitting} based on a local coordinate system which we introduce in subsection \ref{SEC:LocalCoordinate}.
With these forces the active membrane dynamics can be bidirectionally coupled to a surrounding fluid flow which is computed separately using either an overdamped dynamics or a Lattice-Boltzmann method.

\subsection{Local coordinate system}
\label{SEC:LocalCoordinate}

\begin{figure}
	\centering
	\includegraphics[width=.6\textwidth]{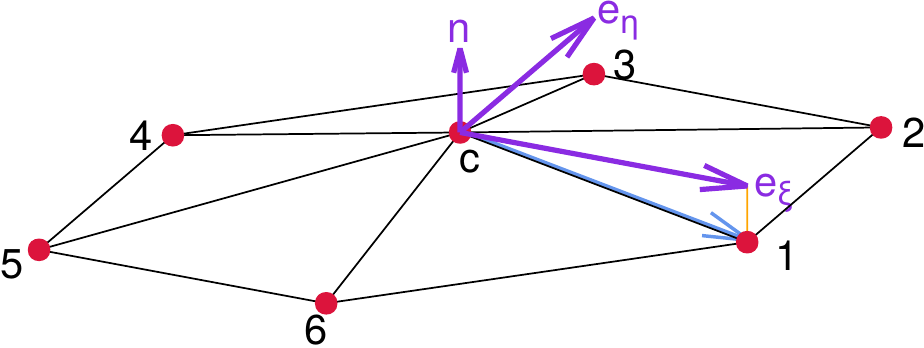}
	\caption{The central node at $\vect{r}_{\text{c}}$ is surrounded by $N$ neighbors.
The discretization is chosen such that all nodes on the cylindrical mesh in figure \ref{FIG:delta3D} have 6 neighbors.
For the ellipsoid in figure \ref{FIG:FlowEllipsoid}, topology requires at least 12 triangles with $N=5$ while all others have $N=6$.
		From the surrounding triangles a local normal vector $\vect{n}$ at $\vect{r}_{\text{c}}$ can be computed.
		Together with the position of one neighbor local in-plane coordinate vectors $\vect{e}_\xi, \vect{e}_\eta$ can be constructed.
		Applying the first step of Gram-Schmidt orthonormalization leads to $\vect{e}_\xi$ and the cross product of $\vect{n}$ and $\vect{e}_\xi$ to $\vect{e}_\eta$.}
	\label{FIG:neighborhood}
\end{figure}

As sketched in figure \ref{FIG:neighborhood} each membrane node $\vect{r}_c$ has a neighborhood consisting of $N$ nodes, which are labeled in an ordered fashion around the central node.
The choice of the starting node is arbitrary initially, but has to be retained through the simulation.
In figure \ref{FIG:referenceNodes} we provide evidence that the choice of the starting node does not affect simulation results. 
At the site of the central node a local coordinate system ($\vect{e}_\xi, \vect{e}_\eta, \vect{n}$) can be defined, where we denote the in-plane coordinate vectors ($\vect{e}_\alpha$ ($\alpha=\xi, \eta$) in the appendix) by $\vect{e}_\xi$ and $\vect{e}_\eta$. 
We determine the local unit normal vector $\vect{n}$ by the mean weight by angle of the normal vectors on the surrounding triangles \cite{MeyerDiscretedifferentialgeometryoperators2003}.
The first in-plane vector $\vect{e}_\xi$ is calculated using the vector from the central node to the first neighbor $\vect{x}_1 = \vect{r}_1 - \vect{r}_c$.
The first step of the Gram-Schmidt orthogonalization is applied and the vector is normalized
\begin{equation}
	\vect{e}_\xi = \frac{ \vect{x}_1 - \left(\vect{x}_1 \cdot \vect{n}\right) \vect{n} }{ \vert \vect{x}_1 - \left(\vect{x}_1 \cdot \vect{n}\right) \vect{n} \vert }.
	\label{EQ:unitXi}
\end{equation}
The second in-plane coordinate vector is calculated by the cross-product
\begin{equation}
	\vect{e}_\eta = \frac{\vect{n} \times \vect{e}_\xi}{\vert \vect{n} \times \vect{e}_\xi \vert}.
		\label{EQ:unitEta}
\end{equation}
Using this method we can assign to every node at every time step a unique coordinate system which is adapted to the local deformed membrane geometry.  
We denote coordinates along $\vect{e}_\xi$ and $\vect{e}_\eta$ by $\xi$ and $\eta$, respectively.
In a figurative sense, we co-move with a cytoskeletal filament initially positioned at $\vect{r}_c$ pointing towards $\vect{x}_1$.

\subsection{Parabolic fitting}
\label{SEC:ParabolicFitting}

To obtain local geometrical quantities such as the curvature tensor on the deformed membrane we perform a parabolic fitting procedure based on the local coordinate system derived in the previous section.
An arbitrary point $\bar{\vect{r}}$ in the neighborhood of the central node $\vect{r}_c$ can be expressed by a Taylor expansion around $\vect{r}_c$ up to the second order.
For the $i$-th component of the vector $\bar{\vect{r}}$ with $i=x,y,z$ we obtain
\begin{equation}
	\bar{r}_{i}(\xi,\eta) \approx r_{i}^c + A_{i} \xi + B_{i} \eta + \frac{1}{2} \left( C_{i} \xi^2 + D_{i} \eta^2 + 2E_{i} \xi\eta \right), \\
	\label{EQ:parabolicExpansion}
\end{equation}
with $\xi$, $\eta$ being the coordinates along $\vect{e}_\xi$ and $\vect{e}_\eta$, respectively.
Using this expression we can apply a parabolic fitting procedure (see also \cite{guckenberger_bending_2016, farutin_3d_2014} where a similar procedure has been used to compute passive bending forces) considering all $N$ neighboring nodes with a squared deviation from the fitted surface
\begin{equation}
\chi_i^2 = \sum_{\nu=1}^N \left( r_{i}^{\nu} - \bar{r}_{i}^{\nu} \right)^2,
\label{EQ:chiSquared}
\end{equation}
with $i=x,y,z$.
By minimizing the $\chi_i^2$ we obtain the coefficients $A_i-E_i$. 
Using equation \eqref{EQ:parabolicExpansion} we are able to analytically calculate the derivative of $\chi_i^2$ with respect to the coefficients $A_i-E_i$. 
The derivative of $\chi_i^2$ being zero in case of minimization then leads to a linear equation system for the $A_i-E_i$.
This linear equation system is solved numerically in the simulation using LU decomposition.
The paraboloid fitted to the neighborhood around $\vect{r}_c$ provides a good approximation of the local curvature for typical cell shapes \cite{guckenberger_bending_2016}.
By construction, the fitting coefficients equal the derivatives of the membrane parametrization vector $\bar{\vect{r}}$ with respect to local coordinates at the site of the central node
\begin{equation}
	A_i = \bar{r}_{i,\xi} ~~~ B_i = \bar{r}_{i,\eta} ~~~ C_i = \bar{r}_{i, \xi\xi} ~~~ D_i = \bar{r}_{i,\eta\eta} ~~~ E_i = \bar{r}_{i,\xi\eta}
\end{equation}
or, equivalently,
\begin{equation}
	\vect{A} = \bar{\vect{r}}_{,\xi}\vert_{\vect{r}_c} ~~~ \vect{B} = \bar{\vect{r}}_{,\eta}\vert_{\vect{r}_c} ~~~ \vect{C} = \bar{\vect{r}}_{,\xi\xi}\vert_{\vect{r}_c} ~~~ \vect{D} = \bar{\vect{r}}_{,\eta\eta}\vert_{\vect{r}_c} ~~~ \vect{E} = \bar{\vect{r}}_{,\xi\eta}\vert_{\vect{r}_c}.
\end{equation}
Thus, we are able to calculate geometrical quantities, as defined in the appendix, at the site of the central node in local coordinates with $\alphaLC,\betaLC,\gammaLC = \xi, \eta$, such as the metric tensor
\begin{equation}
	g_{\alphaLC\betaLC}  = \begin{pmatrix}
	  \vect{A}\cdot \vect{A} & \vect{A}\cdot \vect{B}\\
	  \vect{B}\cdot \vect{A} & \vect{B}\cdot \vect{B}
	\end{pmatrix},
\end{equation}
the curvature tensor
\begin{equation}
	C_{\alphaLC\betaLC} = \begin{pmatrix}
		- \vect{C} \cdot \vect{n} & - \vect{E} \cdot \vect{n}\\
		- \vect{E} \cdot \vect{n} & - \vect{D} \cdot \vect{n}.
	\end{pmatrix},
\end{equation}
and the derivatives of the metric tensor, e.g.,
\begin{equation}
  g_{\xi\xi,\xi} = \left(\vect{A}\cdot\vect{A}\right)_{,\xi} = 2 \vect{A} \cdot \vect{C},
\end{equation}
which are necessary for the calculation of the Christoffel symbols.
The tensor $g^{\alphaLC\betaLC}$ is obtained by inverting the metric $g_{\alphaLC\betaLC}$ since $g_{\alphaLC\gammaLC}g^{\gammaLC\betaLC} = \delta_{\alphaLC}^{\betaLC}$.
Thus, we can further calculate $C_{\alphaLC}^{~\betaLC} = C_{\alphaLC\gammaLC}g^{\gammaLC\betaLC}$ and $C^{\alphaLC\betaLC} = g^{\alphaLC\gammaLC}C_{\gammaLC}^{~\betaLC}$.
We note that this procedure can be used without any restriction on the deformed membrane as well.
Correspondingly, we obtain the metric tensor $g^\prime_{\alphaLC\betaLC}$, the curvature tensor $C^\prime_{\alphaLC\betaLC}$, and the Christoffel symbols on the deformed surface.

Equation \eqref{EQ:activeForceTang} for the in-plane, active force includes a partial derivative $t_{\ia~,\alpha}^{\alpha\beta}$, which is calculated using another parabolic fitting procedure.
As done in equation \eqref{EQ:parabolicExpansion} for the position in local coordinates we can expand the active in-plane surface stress $t^{\ab}$ around the components of the central node $t^{c\ab}_\ia$
\begin{equation}
  \bar{t}^{\ab}_\ia (\xi, \eta) =  t^{c\ab}_\ia + A_\ia^\ab \xi + B_\ia^\ab \eta + \frac{1}{2} \left( C_\ia^\ab \xi^2 + D_\ia^\ab \eta^2 + 2 E_\ia^\ab \xi \eta \right)
\end{equation}
Corresponding to equation \eqref{EQ:chiSquared} we consider here the squared deviation from the expanded active stress $\bar{t}_\ia^{\tilde{\alpha}\tilde{\beta}}$
\begin{equation}
	\chi_\ia^2 = \sum\limits_{\nu=1}^{N} ( t_\ia^{\tilde{\alpha}\tilde{\beta}\nu} -  \bar{t}_\ia^{\tilde{\alpha}\tilde{\beta}\nu} )^2
	\label{EQ:parabolicFitActiveStress}
\end{equation}
with $t_\ia^{\tilde{\alpha}\tilde{\beta}\nu}$ being one component of the active in-plane surface stress of the $\nu$-th neighboring node in local coordinates.
Thus, the corresponding fitting coefficients $A_\ia$ and $B_\ia$ represent the partial derivative of the active in-plane surface stress component with respect to the coordinate $\xi$ and $\eta$, respectively, whereas the coefficients $C_\ia \-- E_\ia$ represent the second derivatives.
We note that the matrix for the fit is the same as for the position.
We perform one fitting procedure for each component, i.e., $\tilde{\alpha} = \xi, \eta$ and $\tilde{\beta} = \xi, \eta$. 

With the method proposed here, we are able to calculate the co-variant derivative, which consists of a partial derivative and Christoffel symbols, and perform the contraction of active in-plane surface stress tensor and curvature tensor in the local coordinate system to determine the traction jump given by equations~(\ref{EQ:activeForceTang}) and (\ref{EQ:activeForceNorm}), which we repeat here for the deformed membrane in the local coordinate system
\begin{align}
f_\ia^\betaLC  &= t^{\alphaLC\betaLC}_{\ia~,\alphaLC} + \Gamma_{\alphaLC\gammaLC}^{\prime\alphaLC} t_\ia^{\gammaLC\betaLC} + \Gamma_{\alphaLC\gammaLC}^{\prime\betaLC} t_\ia^{\alphaLC\gammaLC}\label{EQ:activeForceTangDeformed}\\
f_\ia^n &= - C^{\prime}_{\alphaLC\betaLC} t_\ia^{\alphaLC\betaLC},
\label{EQ:activeForceNormDeformed}
\end{align}
in case of vanishing active moments and vanishing transverse shear stress.
We note that the in-plane surface stress is the analogue of the Cauchy stress tensor in 3D \cite{berthoumieux_active_2014,barthes-biesel_motion_2016} and thus acting on the deformed membrane \cite{chandrasekharaiah_continuum_2014}.
Equation \eqref{EQ:activeForceTangDeformed} and \eqref{EQ:activeForceNormDeformed} give the discrete forces (per area) on one membrane node.
These forces then enter the fluid solver as illustrated in section \ref{SEC:IBM} or are used for the relaxation dynamics in the overdamped limit (see appendix \ref{SEC:method}).

To convert the traction jump from equations~(\ref{EQ:activeForceTangDeformed}) and (\ref{EQ:activeForceNormDeformed}) into a nodal force we use Meyer's mixed area approach \cite{guckenberger_theory_2017,MeyerDiscretedifferentialgeometryoperators2003}: in the case of non-obtuse triangles the area is calculated using Voronoi area $A_{\text{Voronoi}}$ defined by
\begin{equation}
	A_{\text{Voronoi}} = \frac{1}{8} \sum\limits_{\nu=1}^N \left( \cot(\alpha_{\nu}) + \cot(\beta_\nu) \right) \lvert \vect{x}_\nu \rvert
\end{equation}
with the angles $\alpha$ and $\beta$ opposite to the edge connecting the central node and the $\nu$-th neighbor node within the adjacent triangles and in the case of obtuse triangles the midpoint of $\vect{x}_\nu$ that is opposite of the obtuse angle is chosen instead of the circumcenter point for each triangle.

\subsection{Specification of active in-plane surface stress}
\label{SEC:specificationActive}

Active stresses in membranes are often generated by ATP-fueled molecular motors "walking" along cross-linked rod-like structures such as actin filaments or microtubules.
As a direct consequence, active stresses are often anisotropic with the direction of contractile/extensile stresses specified in a material frame moving and deforming along with the membrane itself.
This naturally leads to a convenient specification of the active in-plane surface stress tensor $t_\ia^{\alpha\beta}$ in the local coordinate system introduced in section~\ref{SEC:LocalCoordinate}.
Since the labeling of the neighbors around each node remains unchanged throughout the simulation, the distance vector $\vect{x}_1 = \vect{r}_1 - \vect{r}_c$ represents a material vector.
Its normalized in-plane counterpart $\vect{e}_\xi$, given in equation~(\ref{EQ:unitXi}), together with $\vect{e}_\eta$, given in equation~(\ref{EQ:unitEta}), constitute the associated material frame.
Imagining one cytoskeletal filament for an illustrative picture, the anchoring position of the filament is tracked by the node position $\vect{r}_c$, while the orientation of the filament is tracked by the fixed choice of the first neighbor $\vect{r}_1$.
We again note that the choice of the starting node for the labeling, here 1, does not affect simulation results, as shown in figure \ref{FIG:referenceNodes}.

The active in-plane surface stress tensor $t_\ia^{\alpha\beta}$ itself is not computed by our method but needs to be specified as an input quantity according to a corresponding constitutive law for the surface stress \cite{berthoumieux_active_2014, salbreux_mechanics_2017-1}.
Our algorithm allows for an arbitrary choice of active in-plane surface stress, subject to condition (\ref{EQ:forceBalanceMN}) for vanishing active moments, including spatially heterogeneous, anisotropic or time-dependent stresses.
For the latter, coupling to a convection-diffusion model of active substances such as ATP or myosin (with the magnitude of $t_\ia^{\alpha\beta}$ proportional to local ATP/myosin concentration) is methodologically possible.
Thus, if the concentration field of ATP/myosin is solved/prescribed on the membrane, e.g. by discretizing the convection-diffusion equation, the active stress can be calculated from the local concentration.
The corresponding active forces and their coupling to the surrounding fluid dynamics are then straightforwardly achieved by the present algorithm, which allows for a spatially varying active stress.
In the present work, we consider only temporally constant active stresses.
Active stresses can thus conveniently be specified in the initial configuration of the membrane.
For this, we first choose an intuitive coordinate system ($\vect{e}_1$,$\vect{e}_2$) appropriate for the initial shape of the undeformed cell membrane.
The active in-plane surface stresses in this coordinate system are of the form
\begin{equation}
     t_{\ia\alpha}^{~~\beta} = \begin{pmatrix}
        t_{\ia 1}^{~~1} &t_{\ia 1}^{~~2} \\
        t_{\ia 2}^{~~1} & t_{\ia 2}^{~~2}
    \end{pmatrix}
    \label{EQ:activeInitial}
\end{equation}
where the mixed form with upper and lower index is chosen such that the right-hand side contains physical material-specific constants with dimensions of N/m \cite{berthoumieux_active_2014}.
Because of the dynamical deformation of the membrane, the active in-plane surface stress of equation~\eqref{EQ:activeInitial} needs to be mapped into the local coordinate system at each node.
This can be achieved by mapping the coordinate system ($\vect{e}_1$,$\vect{e}_2$) to the local coordinate system ($\vect{e}_\xi$,$\vect{e}_\eta$).

In many situations, the intuitive coordinate system ($\vect{e}_1$,$\vect{e}_2$) will correspond to cylindrical coordinates or spherical coordinates, e.g.,~for a rounded cell during mitosis \cite{salbreux_actin_2012}.
By construction of the intuitive in-plane coordinate system and the local coordinate system both normal vectors, e.g.,  $\vect{e}_r$ for a cylinder or a sphere and the local $\vect{n}$ are equal initially. 
Thus, the initial in-plane coordinate system ($\vect{e}_1$,$\vect{e}_2$) can be converted directly into the local coordinate system ($\vect{e}_\xi$,$\vect{e}_\eta$) on every node individually
\begin{equation}
 t_\ia^{\alphaLC\betaLC}  = \mathfrak{D} ~ t_\ia^{\ab} ~ \mathfrak{D}^{-1},
\label{EQ:rotation}
\end{equation}
with $\mathfrak{D}$ being a rotation matrix around the local unit normal vector $\vect{n}$, with $\alpha=1,2$ and $\alphaLC=\xi,\eta$.
We thus obtain the active in-plane surface stress along the local coordinate vectors $\vect{e}_\xi$ and $\vect{e}_\eta$.
The rotation in equation~(\ref{EQ:rotation}) is performed once at the beginning of a simulation.
Since the local coordinate system $\vect{e}_\xi$, $\vect{e}_\eta$ is co-moving with the membrane material, the active in-plane surface stress tensor $t_\ia^{\alphaLC\betaLC}$, expressed in these coordinates, does not change over time.

The parabolic fitting procedure in equation \eqref{EQ:parabolicFitActiveStress} requires the difference in active stress between neighboring nodes.
The differences in active stress are calculated in the intuitive coordinate system, in which the active stresses are prescribed.
Because active stresses do not change in time, calculation of the differences can be done also once at the beginning of the simulation.
Projection into the local coordinate system for each node is done in the same way as for active stresses in equation \eqref{EQ:rotation}.

Along a certain direction, the cytoskeletal filaments may tend to contract or to expand, respectively, resulting in a contractile or extensile active in-plane surface stress \cite{needleman_active_2017}. 
The contractile or extensile nature of the cytoskeletal filaments manifests itself in the sign of the active stress, namely a positive (negative) stress corresponds to a contractile (extensile) nature. 
In active matter consisting of cytoskeletal filaments, active stresses often possess different signs in different directions.
Imagining two anti-parallel polar filaments that are cross-linked by a motor protein walking in opposite directions on both filaments, a relative extensile shift of both filaments together with a lateral contraction occurs.

\section{Validation}
\label{SEC:validation}

In this section we provide an in-depth validation of our algorithm to compute the dynamics of active membranes embedded in a 3D fluid.
For this, we first compute the deformation obtained when an initially unstressed cylindrical membrane is subjected to a localized perturbation due to active stresses.
Our numerical results are in excellent agreement with analytical predictions by \citeauthor{berthoumieux_active_2014}  \cite{berthoumieux_active_2014} which were obtained using a Green's function approach in the limit of small deformations.
Next, we apply homogeneous active stresses again to an initially cylindrical membrane.
In agreement with the analytical predictions of \cite{berthoumieux_active_2014}, we observe two kinds of axisymmetric instabilities.
Going beyond the axisymmetric calculations of \cite{berthoumieux_active_2014}, our numerical method then predicts the existence of a third non-axisymmetric instability which, to the best of our knowledge, has not been observed thus far.

To account for the dynamics of the surrounding fluid, we employ two qualitatively different approaches.
In the first approach, we use simple overdamped dynamics such that the surrounding fluid acts purely as a viscous frictional damping, see appendix \ref{SEC:method}.
In the second approach, we consider the full fluid dynamics of the surrounding liquid by solving the Navier-Stokes equations using a Lattice-Boltzmann method, see section \ref{SEC:LBM}.
Two-way coupling to the active membrane dynamics is provided by the Immersed-Boundary Method, as detailed in section \ref{SEC:IBM}.

\subsection{Green's function formalism}

We here briefly recall the central analytical results of \cite{berthoumieux_active_2014} which will be used to validate our numerical computations in the subsequent paragraphs.
In \cite{berthoumieux_active_2014} the active in-plane surface stresses have the following form 
\begin{equation}
    t_{\ia\alpha}^{~\beta} = \begin{pmatrix}
    T_\ia + T_\ia^z \delta (z) & 0 \\
    0 & T_\ia + T_\ia^\phi \delta (z)
  \end{pmatrix}
  \label{EQ:activeBerthoumieux}
\end{equation}
on the initially unperturbed cylinder surface where the local coordinates $\alpha$ and $\beta$ correspond to polar coordinates $z$ and $\phi$, respectively (cf.~\ref{SEC:specificationActive}).
We note that the component $t_{\ia z}^{~z}$ being positive represents a contractile stress along the cylinder axis and $t_{\ia z}^{~z}$ being negative an extensile stress, as seen on the basis of the buckling instability reported for negative stress in ref. \cite{berthoumieux_active_2014}.
A positive $t_{\ia \phi}^{~\phi}$ represents a contractile stress in azimuthal direction, which causes a contraction of the cylinder.
The latter becomes clear by considering the deformation caused by $T_\ia^\phi$ according to the Green's function.
In equation~(\ref{EQ:activeBerthoumieux}) $T_\ia$ represents an isotropic homogeneous background active stress while $T_\ia^z$ and $T_\ia^\phi$ are the amplitudes of point active stresses along each of the two coordinate axes. $\delta(z)$ is the Dirac delta distribution.

As in ref.~\cite{berthoumieux_active_2014}, we focus on the radial deformation $u_r(z)$ of a cylinder with initial radius $R$ resulting from an azimuthal in-plane surface stress $T_\ia^\phi$.
At the end of section \ref{SEC:LocalSmooth} we perform a validation for an axial in-plane surface stress $T_\ia^z$.
\citet{berthoumieux_active_2014} consider a 3D elastic material and perform a projection onto the membrane resulting in the surface stretching modulus $S$ and the bending modulus $B$ as surface elastic parameters, together with the 3D Poisson ratio $\nu$.
The strength of the homogeneous active in-plane stress is measured by the dimensionless number $g=\frac{T_\ia}{S}$ and bending forces are quantified by the relative bending modulus $b=\frac{B}{SR^2}$.
The radial deformation of a cylindrical shell with radius $R$ is then given by the Green's function  $G_{r\phi}(z)$ as
\begin{align}
  \frac{u_r(z)}{R} &= - G_{r\phi}(z) T_\ia^\phi  = - \frac{ T_\ia^\phi}{RS} G(z). \label{EQ:GreensPoint}
\end{align}
An expression for $G(z)$ is given in Fourier space \cite{berthoumieux_active_2014} by
\begin{align}
  G(q) &= \frac{1}{2b(Rq)^4 + (g-2\nu b)(Rq)^2 + 2(1-\nu^2)-g} \label{EQ:GreensFT}
\end{align}
with $G(z)$ being the inverse Fourier transform of $R G(q)$.

In our numerical method, elastic forces are derived from the Skalak and Helfrich energy functionals as defined in equation \eqref{EQ:Skalak} and \eqref{EQ:Helfrich}, respectively, which represent an accurate and widely used description of the elastic properties of biological cell membranes.
In the limit of small deformations we show in section \ref{SEC:Moduli} that the Skalak and Helfrich model, which we use in this study, and the elastic model used by \citet{berthoumieux_active_2014} are related by $S = \frac{2}{3}\kappa_{\text{S}}$ and $B = \frac{1}{2} \kappa_{\text{B}}$.
In the following these relations are used to calculate the Green's function in equation \eqref{EQ:GreensFT} and the corresponding deformation in equation \eqref{EQ:GreensPoint} for comparison to our simulations.
We furthermore set the bulk Poisson ratio $\nu=\frac{1}{2}$, corresponding to a 3D incompressible material, and the Skalak parameter $C=1$.
Then the Skalak law becomes equivalent to the Neo-Hookean membrane law which is built to model a membrane made of a 3D incompressible material \cite{barthes-biesel_motion_2016}.
For the shear modulus $S$ we show in appendix \ref{SEC:Moduli} that the shear related in-plane surface stresses determining the Green's function agree between the Skalak law and the model used by ref.~\cite{berthoumieux_active_2014} in the limit of small deformation.
Although we can relate the parameter of Helfrich model to the model used in \citet{berthoumieux_active_2014} by  $B = \frac{1}{2} \kappa_{\text{B}}$ as illustrated in the appendix, the Helfrich model alters the in-plane surface stresses, which in turn alter the Green's function.
However, since we consider the limit $B\rightarrow0$ this has no effect for the Green's function used for validation of our simulations.

\subsection{Singular active perturbation}

\begin{figure}
	\centering
	a)
	\begin{minipage}{0.29\textwidth}
		\includegraphics[width=\textwidth]{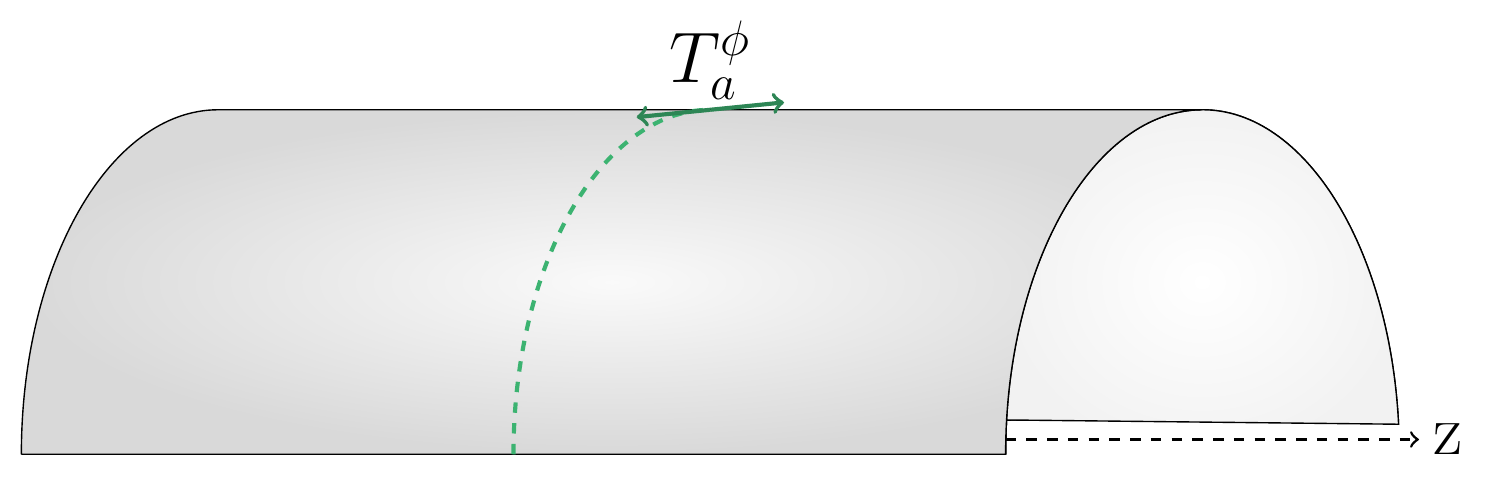}
	\end{minipage}
	b)
	\begin{minipage}{0.29\textwidth}
		\includegraphics[width=\textwidth]{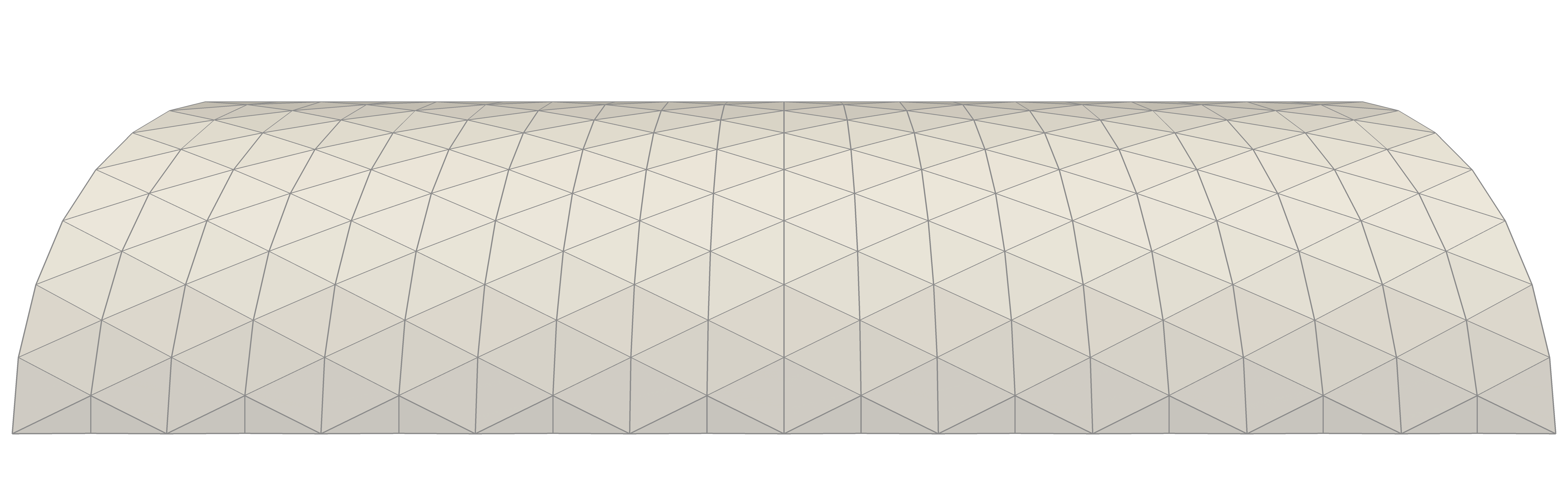}
	\end{minipage}
	c)
	\begin{minipage}{0.29\textwidth}
		\includegraphics[width=\textwidth]{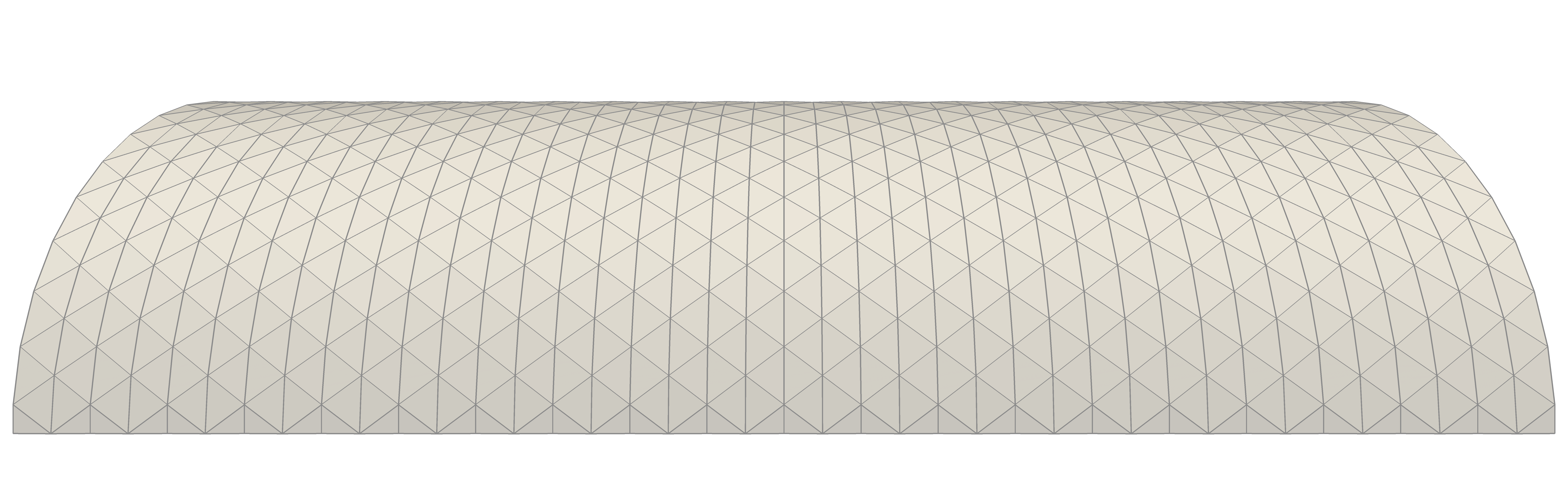}
	\end{minipage}	
	
	\caption{a) We consider a local increase in active in-plane surface stress of a cylindrical membrane. 
	b) and c) Membrane meshes as in main text with two different resolutions a) $\Delta z = 0.2$, $\Delta \phi=0.2$ and b) $\Delta z = 0.1$, $\Delta \phi=0.15$, respectively.
	}
	\label{FIG:delta3D}
\end{figure}

We start by simulations of overdamped dynamics of an initially stress-free cylindrical shell subjected to a local increase in active in-plane surface stress.
The active in-plane surface stress from equation~(\ref{EQ:activeBerthoumieux}) simplifies to
\begin{equation}
t_{\ia\alpha}^{~\beta}  = 
	\begin{pmatrix}
		0 & 0 \\
		0 & T_\ia^\phi
	\end{pmatrix} \delta (z),
\end{equation}
as sketched in figure \ref{FIG:delta3D}~a).
Simulations are carried out with $R=1$, $\kappa_{\text{S}}=1$, $C=1$, $\kappa_{\text{B}} = 0.001$, and $T_\ia^\phi = -0.01$ in simulation units.
We employ two different triangulations of the cylindrical shape which are shown in \ref{FIG:delta3D}~b) and c).
In the coarse mesh, the axial distance between rings of nodes is $\Delta z = 0.2$ and the azimuthal spacing is about $\Delta\phi=0.2$ radians. 
The finer mesh uses an axial spacing of $\Delta z = 0.1$ and an azimuthal spacing of about $\Delta\phi=0.15$ radians.

\begin{figure}
	\centering
	\begin{minipage}{0.45\textwidth}
		\includegraphics[width=\textwidth]{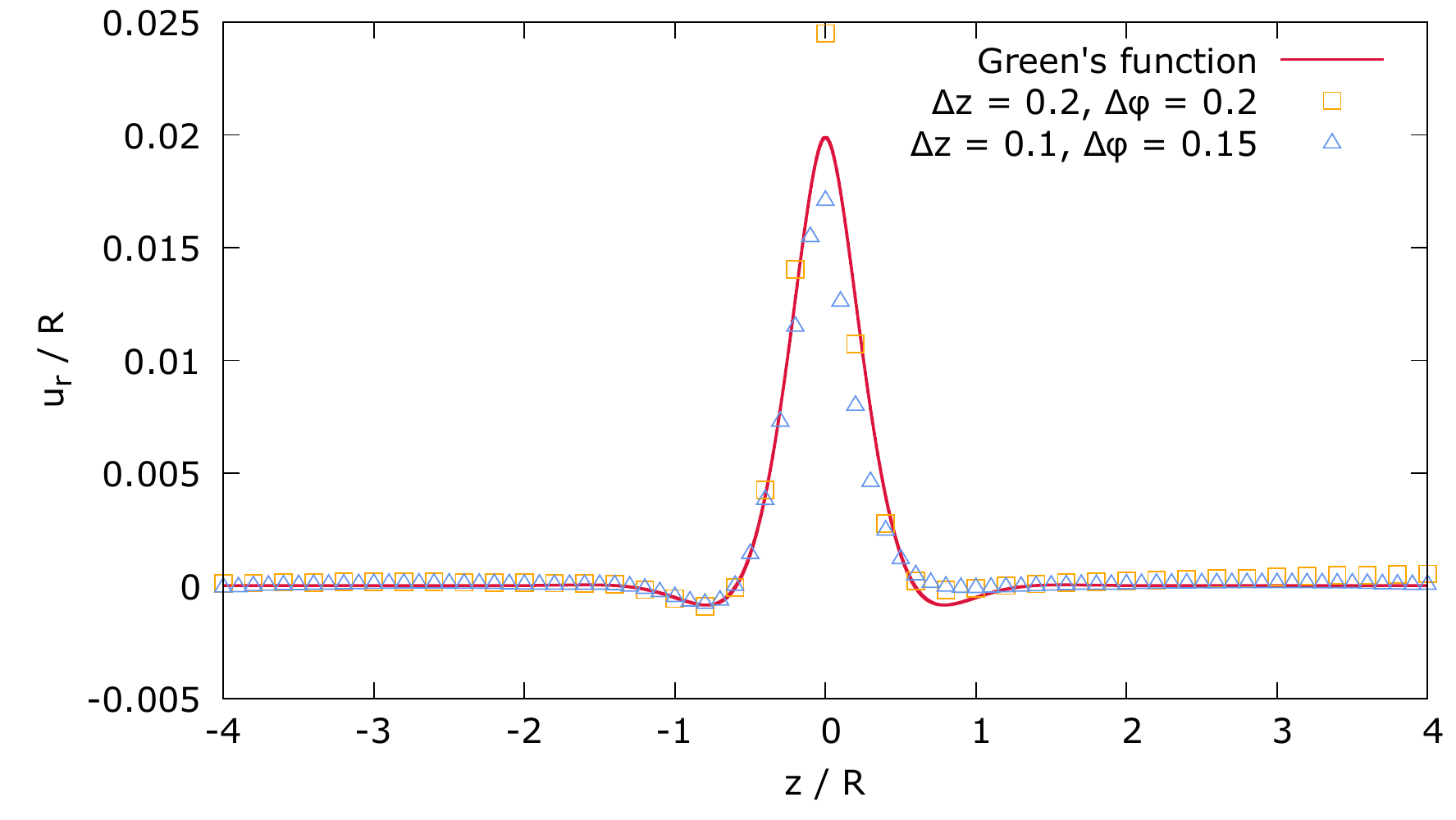}
	\end{minipage}
	\caption{
	The final deformation resulting from three dimensional simulations of a singular active stress shows the same shape as the analytical prediction of the Green's function.
	However, peak height deviates from the theory.
	This can be attributed to the singular nature of the perturbation which is difficult to resolve numerically.
	Deformations are obtained for the parameter set $T_\ia^\phi=-0.01$, $\kappa_S = 1$, $C=1$, and $\kappa_B = 10^{-3}$.
	}
	\label{FIG:delta3Dresults}
\end{figure}

In figure \ref{FIG:delta3Dresults} we compare the final shape of the shell as observed in 3D simulations for the two different resolutions and compare it to the prediction of the Green's function in equation \eqref{EQ:GreensPoint}.
The analytical Green's function shows a peak in deformation of finite width centered at the site of active in-plane surface stress perturbation ($z=0$) that decays with increasing distance, reaches a shallow minimum at around $z\approx \pm 0.7$ and then approaches zero for $z\to\pm\infty$.
Although for both resolutions the resulting amplitude of the peak in deformation from the simulation is close to the Green's function and both show a similar shape, we observe a significant deviation of the 3D simulation results from the theoretical expectation.
Especially, the simulations cannot reproduce correctly the predicted shallow minima next to the main peak.
\new{We note that the deviations for singular perturbation also appear using Lattice Boltzmann/Immersed Boundary method instead of overdamped dynamics (results not shown).}

This observed deviation can be explained by the idealized singular nature of the active in-plane surface stress which is impossible to accurately reproduce on a discretized membrane shape.
Along the cylinder axis, i.e., along $z$-direction, only one ring of nodes with $z=0$ is attributed with finite active in-plane surface stress.
Correspondingly, only these nodes are subjected to active forces and cause the neighboring nodes to move due to the elastic nature of the cylinder.
We note that the central nodes \-- with active in-plane surface stress \-- experience a significantly larger  deformation than their direct, adjacent neighbor nodes, as seen in figure \ref{FIG:delta3Dresults}.
This steep gradient in deformation resulting in locally very large curvature cannot be completely resolved by the parabolic fit.
Thus, the procedure fails to resolve the deformation caused by a point perturbation, although obtained deformations are similar in shape and nearly match the amplitude of the Green's function.
Given that in real applications, all perturbations can be expected to be non-singular, we proceed to investigate the behavior of our algorithm for spatially smooth active stresses.

\subsection{Localized smooth perturbation}
\label{SEC:LocalSmooth}

As a prototypical smooth distribution, we choose a Gaussian distributed active in-plane surface stress of the form
\begin{equation}
t_{\ia\alpha}^{~\beta}  = 
	\begin{pmatrix}
		0 & 0 \\
		0 & T_\ia^\phi
	\end{pmatrix}  \exp\left( -\frac{z^2}{R^2} \right),
\end{equation}
where $T_\ia^\phi$ is again a constant amplitude.
For this distribution, the predicted deformation can be obtained by superposing a distribution of Green's functions (equation (\ref{EQ:GreensPoint})) leading to the deformation
\begin{equation}
  \frac{u_r (z)}{R} = - \int\limits_{-\infty}^{\infty} G_{r\phi}(z-z^{\prime}) T_\ia^{\phi} \exp( -z^{\prime2}/R^2 ) \dInt{z^{\prime}}.
  \label{EQ:defFromGreensPhi}
\end{equation}
We use the Green's function in Fourier space given in equation~(\ref{EQ:GreensFT}) and the Fourier transform of the Gaussian
\begin{equation*}
  \int\limits_{-\infty}^{\infty} \exp\left( -z^2/R^2 \right) \exp(-iqz) \text{d}\left(\frac{z}{R}\right) = \sqrt{\pi} \exp(-\frac{1}{4}(Rq)^2).
\end{equation*}
Using the convolution theorem in Fourier space and transforming back to real space leads to
\begin{equation*}
	\frac{u_r(z)}{R} = -\frac{1}{2\pi} \frac{R}{S} \int\limits_{-\infty}^{\infty}   \frac{\sqrt{\pi} T_\ia^\phi \exp(-\frac{1}{4}(Rq)^2)}{2b(Rq)^4 - 2\nu b (Rq)^2 + 2(1-\nu^2)} \exp(iqz) \text{d}q.
\end{equation*}
The integral can be solved analytically in the limit of small bending rigidity $b\ll 1$ (which corresponds well with the chosen simulation parameters) to obtain
\begin{equation}
   \frac{u_r(z)}{R} = -\frac{T_\ia^\phi}{S} \frac{1}{2(1-\nu^2)} \exp(-\frac{z^2}{R^2}).
\label{EQ:analyticalSmooth}
\end{equation}
Alternatively, we can solve the integral in equation~(\ref{EQ:defFromGreensPhi}) numerically, which does not lead to any differences for small bending modulus (results not shown).

\begin{figure}[!h]
	\centering
	a)
	\begin{minipage}{0.45\textwidth}
	  \includegraphics[width=\textwidth]{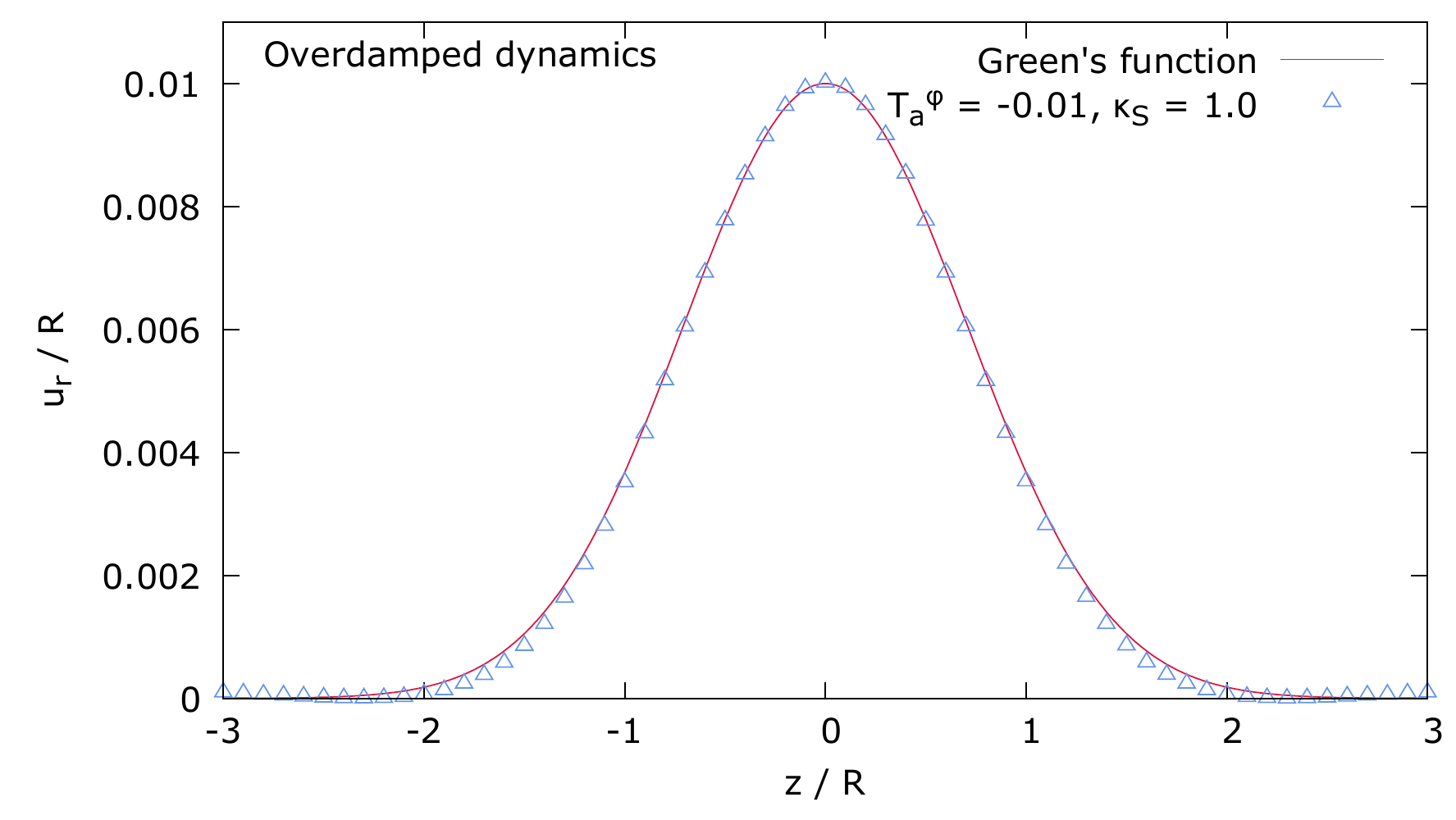}

	\end{minipage}
	b)
	\begin{minipage}{0.45\textwidth}
	  \includegraphics[width=\textwidth]{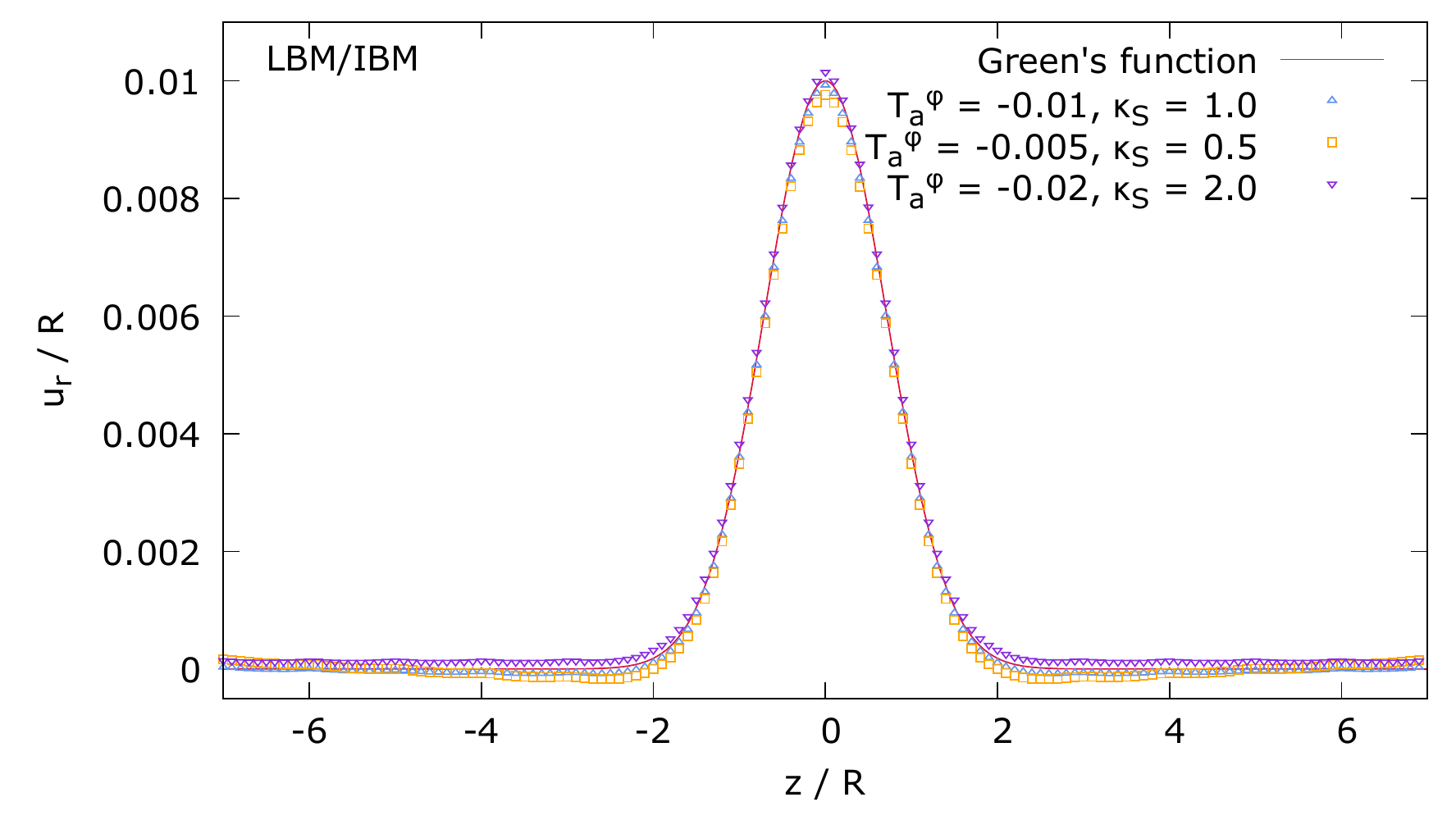}
	\end{minipage}
	
	c)
	\begin{minipage}{0.45\textwidth}
	  \includegraphics[width=\textwidth]{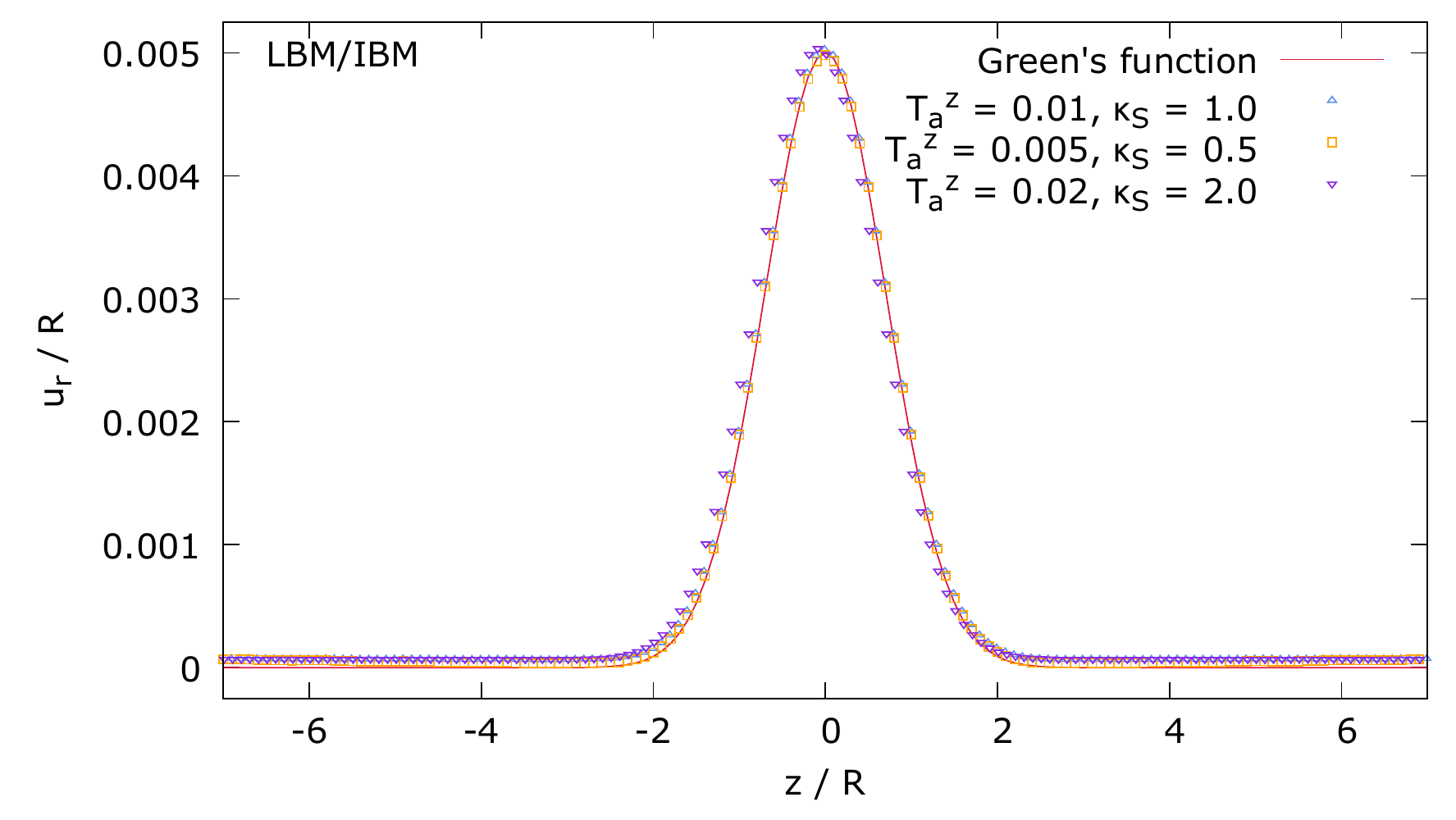}
	\end{minipage}
	d)
	\begin{minipage}{0.45\textwidth}
	  \includegraphics[width=\textwidth]{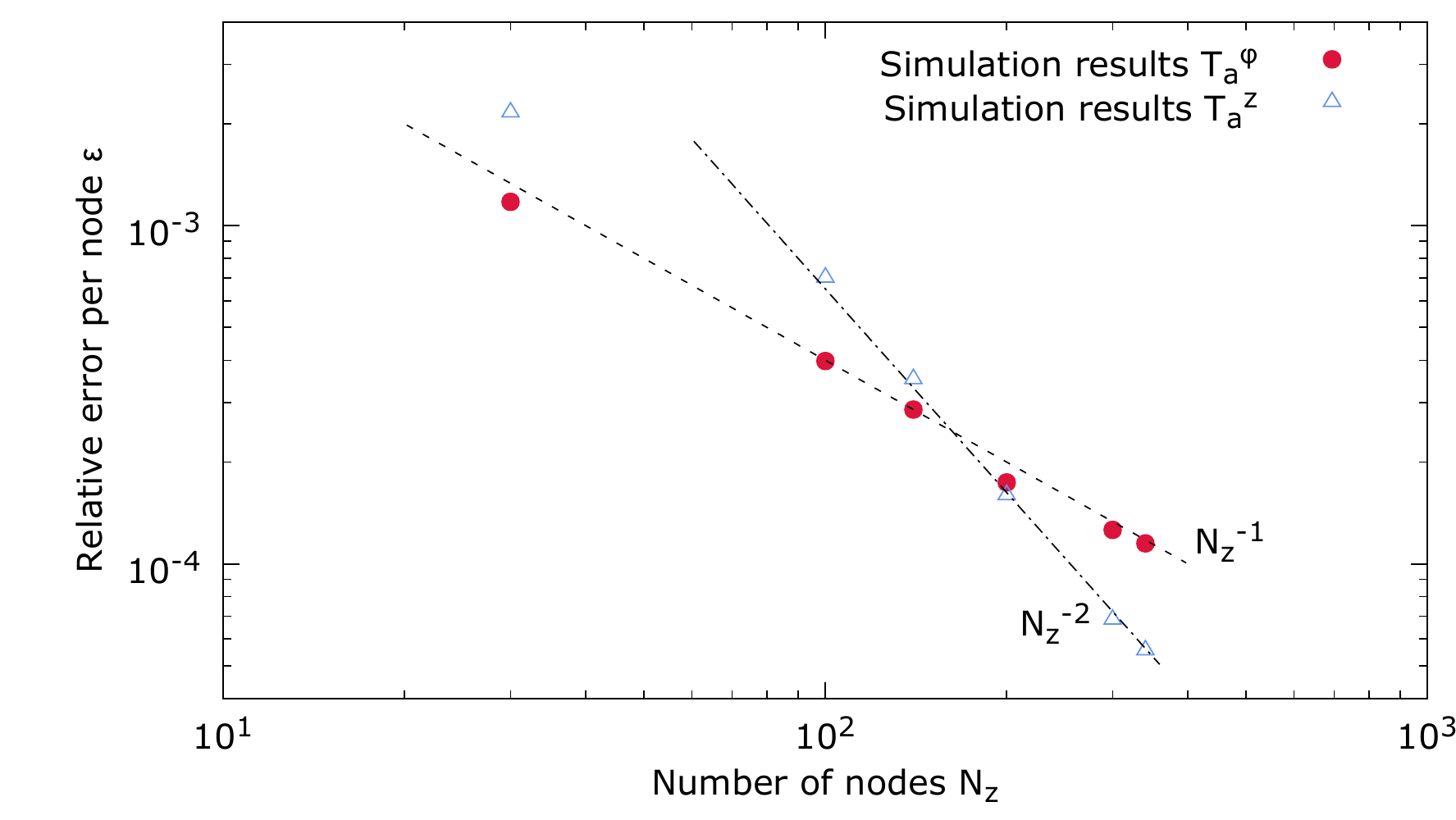}
	\end{minipage}
	\caption{
	a) Comparison of the deformation obtained by the 3D overdamped dynamics with the theoretical expectation for a cylindrical shell with 160 nodes along $z$-direction subjected to a Gaussian distributed active in-plane surface stress with $\kappa_S = 1$, $\kappa_B=10^{-5}$, and $T_\ia^\phi=-0.01$.
	3D simulations are in very good agreement with the theory.
	b) Comparison of the deformation obtained with LBM/IBM simulations for $C=1$ and $\kappa_B \approx 10^{-4}$ with the theoretical prediction for the same setup with perturbation in $\phi$-stress $T_\ia^\phi$ and c) with perturbation in $z$-stress $T_\ia^z$.
	d) Increasing resolution of both membrane and fluid mesh show convergence of the relative error per node for the parameter set of b) for a perturbation in $T_\ia^\phi$ (red dots) as well as for the parameter set of c) for a perturbation in $T_\ia^z$ (blue triangles).
	\new{The error decreases proportional to $N_z^{-1}$ and proportional to $N_z^{-2}$, respectively.}
	In b) \--- c) simulations are done for 300 nodes in $z$-direction. 
	}
	\label{FIG:bellShaped}
\end{figure}

In figure \ref{FIG:bellShaped}~a) we compare the 3D simulation results using overdamped dynamics for parameters $R=1$, $\kappa_{\text{S}} = 1$, $C=1$ , $\kappa_{\text{B}} = 10^{-5}$, and $T_\ia^\phi=-0.01$ to the analytical prediction of equation~(\ref{EQ:analyticalSmooth}).
The Gaussian distribution of active in-plane surface stress leads to a much smoother and broader peak of deformation than the singular perturbation of the previous subsection.
Our simulation results are now in very good agreement with the theoretical prediction.

To go one step further, with applications in mind such as an active membrane in a flowing liquid, we now replace the simple overdamped fluid dynamics with a full Navier-Stokes dynamics solved by the Lattice-Boltzmann method and coupled to the active membrane via the Immersed-Boundary method as described in section \ref{SEC:couplingMemFluid}.
In figure \ref{FIG:bellShaped}~b) we compare simulation results obtained by LBM/IBM to the theoretical Green's function.
Again, our simulations are in very good agreement with the analytical theory.
In figure \ref{FIG:bellShaped}~b) we include three sets of active in-plane surface stress and shear modulus with constant $\kappa_{\text{B}} = 0.00018$ which are chosen such that the ratio of active in-plane surface stress and shear modulus $g$ remains constant. 
Thus, the Green's function predicts identical deformation in all three cases which is indeed observed in our simulations.
 
We now proceed to the validation for a perturbation in $z$-stress, i.e., the active in-plane surface stress takes the form
\begin{equation}
  t_{\ia\alpha}^{~\beta}  = 
	\begin{pmatrix}
		T_\ia^z & 0 \\
		0 & 0
	\end{pmatrix}  \exp( -\frac{z^2}{R^2} ),
\end{equation}
with the constant amplitude $T_\ia^z$.
Corresponding to equation \eqref{EQ:defFromGreensPhi} the deformation can be obtained by \cite{berthoumieux_active_2014}
 \begin{equation}
  \frac{u_r (z)}{R} = \int\limits_{-\infty}^{\infty} G_{rz}(z-z^{\prime}) T_\ia^{z} \exp( -z^{\prime2}/R^2 ) \dInt{z^{\prime}},
\end{equation}
with $G_{rz}(s) = \frac{\nu}{RS} G(z)$ and takes in analogy to equation~(\ref{EQ:analyticalSmooth}) the form
\begin{equation}
   \frac{u_r(z)}{R} = \frac{T_\ia^z}{S} \frac{\nu}{2(1-\nu^2)} \exp(-\frac{z^2}{R^2}).
   \label{EQ:analyticalSmooth_z}
\end{equation}

In figure \ref{FIG:bellShaped} ~c) we compare LBM/IBM simulation results for three different perturbations and shear moduli for $R=1$, $C=1$, and $\kappa_{\text{B}} = 0.00018$ to the theory.
Our simulation results are in very good agreement with the theory and by comparing figure \ref{FIG:bellShaped}~b) and c) we observe half the maximum deformation for $\vert T_\ia^\phi \vert = \vert T_\ia^z \vert$ which is indeed predicted by the theory for $\nu=\frac{1}{2}$.
 
As a further test of algorithm accuracy we now perform a convergence study based on the setup in figure \ref{FIG:bellShaped}~b) as well as in c).
We compare the deformation obtained by simulation $u_r^{\text{sim}}$ and by Green's function $u_r^{\text{Green}}$ by calculating the relative error per node defined as 
\begin{equation}
\epsilon = \frac{1}{N_z} \sqrt{\sum\limits_{z_i} \left( \left(u_r^\text{sim}(z_i) - u_r^{\text{Green}}(z_i)\right)/u_r^{\text{Green}}(0) \right)^2}, 
\end {equation}
where $N_z$ denotes the number of nodes along the cylinder axis and the difference is in relation to the maximal deformation given by the Green's function.
In figure \ref{FIG:bellShaped}~d) we show the relative error per node in dependency of the number of membrane nodes.
With increasing resolution the error per node steadily decreases in both cases.
\new{
The converge rate, however, is different for $T_\ia^\phi$ which decreases with slope $N_z^{-1}$ and $T_\ia^z$ which decreases more quickly with $N_z^{-2}$.
}
This may be due to the fact that the Green's function is derived from linearized equations of motion \cite{berthoumieux_active_2014}, whereas simulations are also valid for larger deformations.
A perturbation in $T_\ia^z$ shows half the maximal deformation as a perturbation in $T_\ia^\phi$ as predicted in equation \eqref{EQ:analyticalSmooth_z} and thus is in better agreement with the Green's function.
The steady decrease in error demonstrates the accuracy of the presented algorithm for active force calculation. 
 
From figure \ref{FIG:bellShaped} we conclude that our algorithm presented in sec.~\ref{SEC:algorithm} together with elastic force calculations gives reliable results for a reasonably smooth distribution of active in-plane surface stress.
The very good accuracy of the predictions is achieved for the simple overdamped dynamics as well as for the substantially more complex and flexible combination of an active membrane with the lattice-Boltzmann/Immersed-Boundary method.

\begin{figure}[!h]
	\centering
	\includegraphics[width=0.75\textwidth]{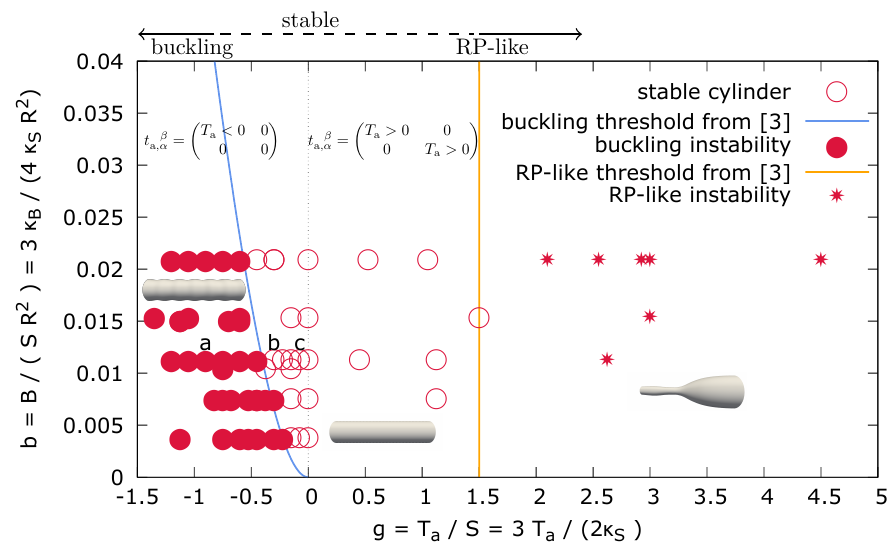}

	\caption{Phase diagram of a cylindrical shell with relative, active in-plane surface stress $g = T_\ia / S = 3 T_\ia/ (2 \kappa_S)$ and relative bending modulus $b = B / (S R^2) = 3 \kappa_B / ( 4 \kappa_S R^2)$ from 3D LBM/IBM simulations in comparison to the theoretically predicted thresholds \cite{berthoumieux_active_2014}.
	On the left of the dotted line we apply a negative active stress, on the right of the dotted line a positive active stress.
	For small negative or positive active stress (around the dotted line) the cylindrical membrane is stable.
	For large negative stresses, a buckling instability is observed.	
	For large positive active in-plane surface stress a Rayleigh-Plateau like instability occurs. 
	The theoretically predicted instability thresholds \cite{berthoumieux_active_2014} and our 3D LBM/IBM simulations are in excellent agreement in both cases.
	Insets illustrate the shape of the shell corresponding to different values of active in-plane surface stress.
	The labels a, b, c refer to figure \ref{FIG:instabilityPhi}.
	}
	\label{FIG:PD_3D}
\end{figure}

\subsection{Homogeneous perturbation: instability diagram}
\label{SEC:PD}

To provide another test of our algorithm we perform simulations of a cylindrical membrane which now is subjected to a homogeneous active in-plane surface stress
\begin{equation}
  t_{\ia\alpha}^{~\beta} = \begin{pmatrix}
    T_\ia & 0 \\
    0 & T_\ia
  \end{pmatrix},
\end{equation}
with $T_\ia = \const$
This situation corresponds to a membrane with constant motor protein activity and isotropic cortex architecture.
We note again that positive $T_\ia$ represents contractile and negative $T_\ia$ extensile stress in $z$- or $\phi$-direction.
Although they did not explicitly compute the deformation for this situation, ref.~\cite{berthoumieux_active_2014} predicts two unstable regions in $g$-$b$-parameter space for which the Green's function could be shown to diverge.
The predicted instability thresholds serve us as a further validation of our simulation method.

By varying both the relative active in-plane surface stress $g = T_\ia / S = 3 T_\ia/ (2 \kappa_S)$ and the relative bending modulus $b = B / (S R^2) = 3 \kappa_B / ( 4 \kappa_S R^2)$ we obtain the phase diagram in figure \ref{FIG:PD_3D} and compare it to the predicted instability thresholds given by \citet{berthoumieux_active_2014}.
On the right, for large $g$, \citeauthor{berthoumieux_active_2014} \cite{berthoumieux_active_2014} predict an instability occurring for $T_\ia > \frac{4}{3}\kappa_S(1-\nu^2)$ shown by the vertical orange line in figure~\ref{FIG:PD_3D}.
Indeed, for simulations in this range we observe an instability with local contraction of the cylinder (see inset of figure \ref{FIG:PD_3D} for a shape illustration).
The threshold obtained by our simulations closely matches the analytically predicted threshold.
This instability is analogous to a Rayleigh-Plateau instability of a liquid jet with the positive, contractile active in-plane surface stress playing the role of the surface tension.
To the left of the threshold a fairly large region is observed in which the initial cylindrical shape remains stable.
For negative $g$ (extensile stress), \cite{berthoumieux_active_2014} predicts a buckling instability when $T_\ia < -2 \sqrt{\frac{\kappa_B\kappa_S}{R^2}}$.
To compare to our simulations, we prescribe an active stress only along the cylinder axis, i.e., $t_{\ia z}^{~z} = T_\ia$ and $t_{\ia\phi}^{~\phi} =0$.
This corresponds to a contracting, cylindrical membrane and the resulting shape beyond the threshold is illustrated in figure \ref{FIG:PD_3D} at the bottom left.
Our simulations agree very well with the predicted instability onset depending on the relative active in-plane surface stress and the relative bending modulus.

\begin{figure}[h!]
	\centering
	\includegraphics[width=\textwidth]{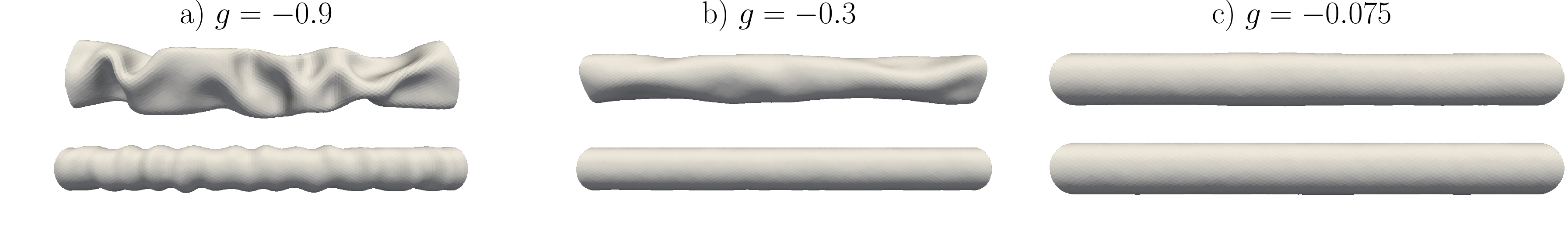}

	\caption{
		Membrane shape for different values of negative active in-plane surface stress for fixed bending modulus $b=0.01125$.
		In the upper row the membrane is subjected to both $z$- and $\phi$-stress, while in the lower row the membrane is subjected to $z$-stress only.
		For $z$-stress only we observe a buckling instability in a).
		However, for isotropic stress we observe an instability introducing non-axisymmetric deformations at intermediate stresses in b) where no buckling instability occurs.
		c) For smaller active stress the cylindrical membrane remains stable in both cases.
		Corresponding points in the phase diagram in figure \ref{FIG:PD_3D} are labeled with a to c.
	}
	\label{FIG:instabilityPhi}
\end{figure}

In addition, we investigate the transition to buckling in more detail and carry out simulations imposing an active tension in azimuthal and axial direction $t_{\ia z}^{~z} = t_{\ia\phi}^{~\phi} = T_a$, which corresponds exactly to the scenario considered by \cite{berthoumieux_active_2014}.
In figure \ref{FIG:instabilityPhi} we compare these simulations (top row) with the ones in figure~\ref{FIG:PD_3D} (bottom row), respectively.
At large (negative) active stress in a) we observe an instability in both simulations, however only the simulation with a purely axial stress clearly corresponds to a buckling instability.
The instability in the top row exhibits a non-axisymmetric character.
For slightly smaller active stress in \ref{FIG:instabilityPhi}~b) the non-axisymmetric instability remains for the isotropic stress, but the membrane becomes stable for $z$-stress only.
Decreasing the active stress further, we observe a stable cylindrical membrane in both cases, as shown in c).

From figure \ref{FIG:instabilityPhi} we conclude that an additional instability is present (not related to buckling) caused by a finite azimuthal stress $t_{\ia\phi}^{~\phi}$.
This additional instability induces non-axisymmetric deformations and thus is not observed in the axisymmetric treatment of ref. \cite{berthoumieux_active_2014} nor in our axisymmetric simulations for which also the onset of the buckling instability is in exact agreement with the analytical prediction as shown in figure S2 of the Supporting Information.

\begin{figure}[!h]
	\centering
	\begin{minipage}{.75\textwidth}
	  \centering
	  \includegraphics[width=\textwidth]{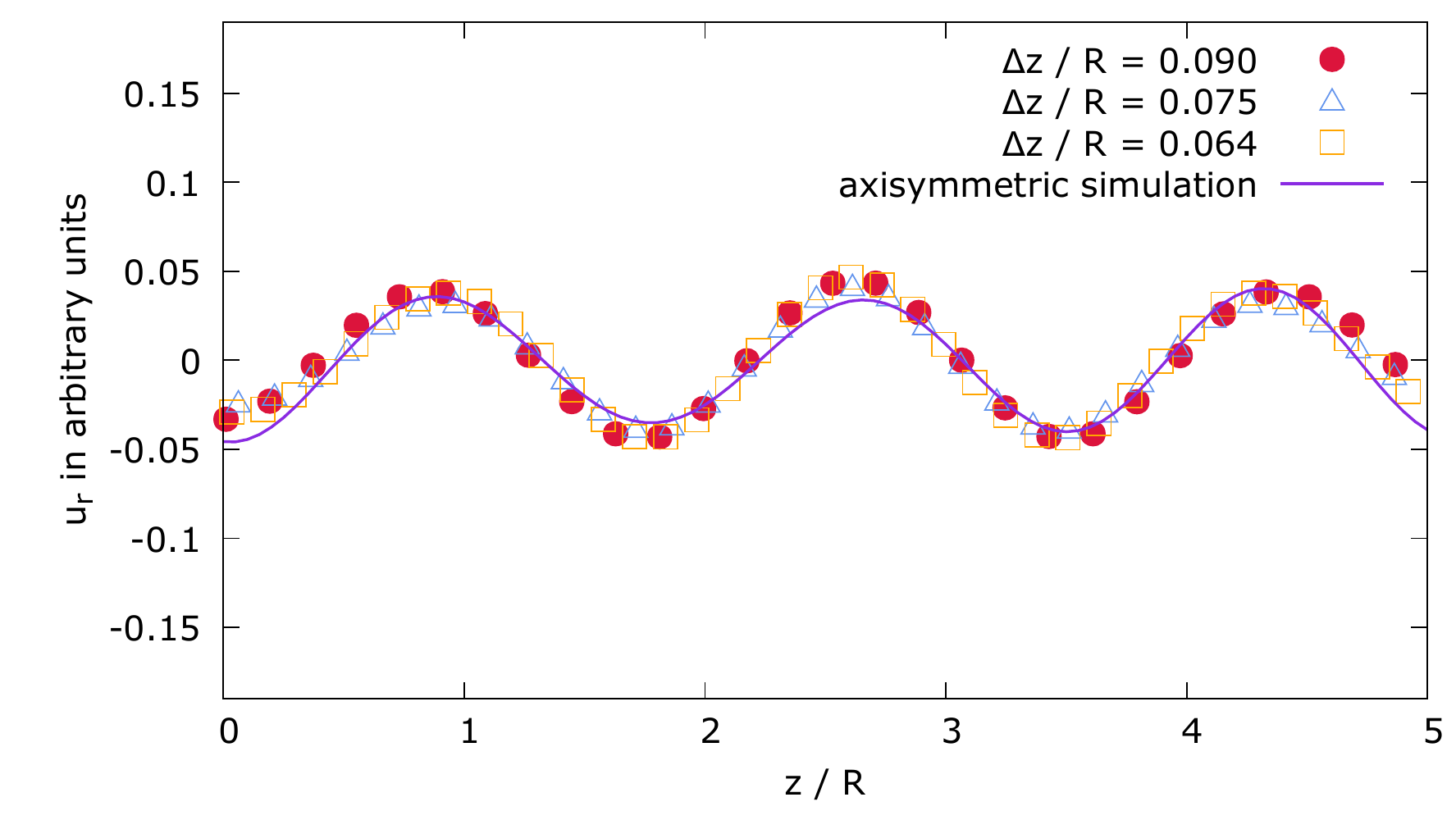}
	\end{minipage}
	\caption{
	Comparison of the deformation obtained for the buckling instability of an initially cylindrical membrane subjected to a homogeneous active in-plane surface stress with $g=-0.75$ and $b=0.01125$.
	We compare the wavelength of the deformation obtained in 3D LBM/IBM simulations with different resolutions $\Delta z / R$ to one obtained by axisymmetric simulation.
	For a sample illustration of the 3D shape we refer to the inset on the left hand side of reference \ref{FIG:PD_3D}.
	All simulations show the same wavelength for the buckling instability.
	}
	\label{FIG:buckling}
\end{figure}

So far, we have validated our algorithm to agree with theoretical predictions on the basis of the Green's function.
To obtain a validation in the non-linear regime beyond the Green's function, we compare our 3D simulations to simulations of an axisymmetric membrane, as detailed in the Supplemental Material \cite{supplemental}.
We do this by comparing the membrane shape in the case of a buckling instability.
In figure \ref{FIG:buckling} we show the deformation obtained from the axisymmetric simulation and compare  it to three simulations using 3D LBM/IBM.
The three 3D simulations are done for different resolutions $\Delta z/R$.
We use the non-dimensional parameters $g=-0.75$ and $b=0.01125$.
On the one hand all three 3D simulations show the same deformation and wavelength and on the other hand they agree in the wavelength with the axisymmetric simulation method.
The wavelength of the buckling instability is about 1.75$R$ for both very different simulation techniques.
In case of small bending elasticity \citet{berthoumieux_active_2014} predict a wavelength at the threshold, where the denominator of the Green's function in eq. \eqref{EQ:GreensFT} becomes zero for finite wave vector $q$, of $\lambda=1.9R$ which is reasonably close to the value observed with simulations. 
The difference may arise from the periodicity of the shell in our simulations and/or from finite bending together with being beyond the threshold.
Nevertheless, the excellent agreement between the axisymmetric and 3D simulations provides strong evidence for the reliability of our algorithm also in the range of large deformations. 

\section{ Model application: cell division in suspending fluid }

In the following we present a first application of our method including fluid flow.
This illustrates the versatility and applicability of our combined LBM/IBM method for active cell membranes.
For this, we consider a dividing ellipsoidal cell.
Except the fact that we employ an elastic rather than a viscous cortex, our setup resembles the situation of cell cytokinesis \cite{salbreux_hydrodynamics_2009,sedzinski_polar_2011,mendes_pinto_force_2013,turlier_furrow_2014,sain_dynamic_2015}.
Cell cytokinesis as part of cell division is a prominent subject of active matter research in biological physics \cite{GrillGrowingstressfulbiophysical2011,green_cytokinesis_2012,pollard_nine_2017}.
Most previous studies, e.g. \cite{turlier_furrow_2014} or \cite{sain_dynamic_2015}, investigated the dynamics of cell cytokinesis for an axisymmetric membrane without considering internal fluid flow. 
Refs.~\cite{li_immersed_2012} and \cite{zhao_modeling_2016} analyzed the flow field inside a dividing cell where the contractile ring is modeled as an additional force using the immersed boundary method and phase field model, respectively.
Ref.~\cite{lee_mathematical_2018} analyzed the flow field by means of the phase field model as well, but considered the cortical ring as shrinking elastic loop.
Here, we consider cell division triggered by active stresses including an external flow leading to a fully 3D asymmetric membrane shape during the division.
We first analyze the flow field dynamically evolving inside the dividing cell surrounded by a quiescent medium in section \ref{SEC:FlowInSpheroid} and then extend this setup by considering a dividing cell in an external shear flow in section \ref{SEC:DividingShearFlow}.

\subsection{Flow field inside a dividing cell}
\label{SEC:FlowInSpheroid}

We consider a prolate ellipsoidal cell of diameter $7~\mu$m and length $14~\mu$m which is endowed with an isotropic active stress $T_\ia = 8\times10^{-5}~\text{N/m}$.
In addition, in an interval of $\Delta\theta \approx \frac{\pi}{12}$ around the equator the active stress in azimuthal direction is increased by a factor of six according to a step function.
The membrane is endowed with shear elasticity $\kappa_S = 5 \cdot 10^{6}$~N/m, $C=1$, and bending elasticity $\kappa_B=2 \cdot 10^{-19}$~Nm which are in the range of typical cell membranes.

\begin{figure}[!h]
	\centering
	\includegraphics[width=\textwidth]{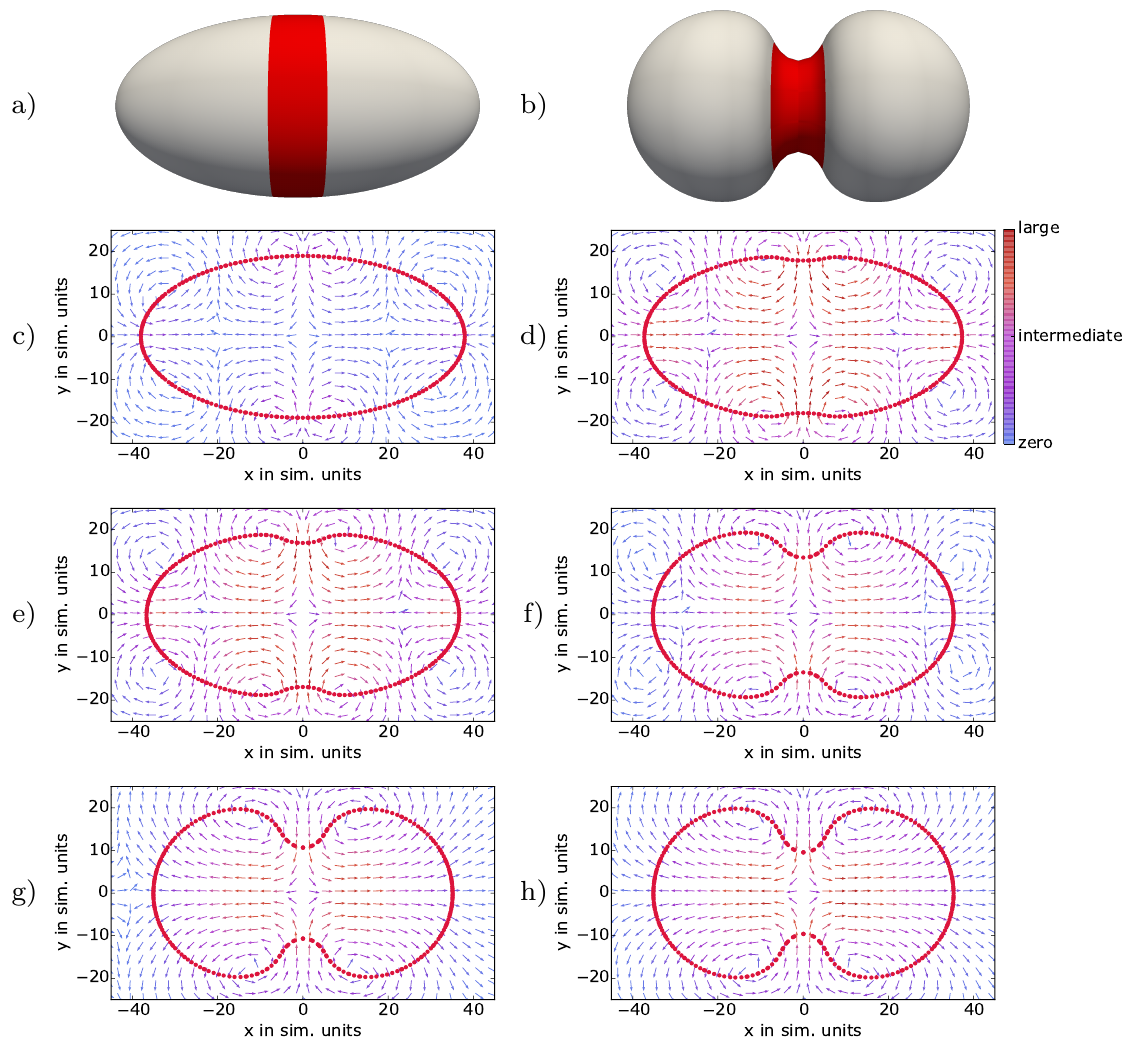}
	\caption{
	a) Similar to cell division an elastic cell membrane, which is subjected to homogeneous active stress and a local increase in azimuthal active stress in the red shaded area, contracts locally as shown in b).
	(c) \-- (h) show the outline of the deforming membrane (red nodes) in the central plane and the developing flow field (arrows) inside the cell over time.
	Eventually, a flow away from the contracting region towards the poles of the cell is observed.
	Arrows indicate the flow direction while the color indicates the flow velocity.	
	\new{ Relative time is $t / t_{\text{max}}$ = a) 0, b) 1, c) 4$\times10^{-4}$, d) 0.08, e) 0.16, f) 0.51, g) 0.86, h) 1.0 .}
	}
	\label{FIG:FlowEllipsoid}
\end{figure}

We present the initial 3D membrane shape in figure \ref{FIG:FlowEllipsoid}~a) and a deformed 3D shape in figure \ref{FIG:FlowEllipsoid}~b).
In both cases we illustrate the region with increased active stress along $\phi$-direction by the red shaded area.
The shape and the developing flow field in the central plane, which includes the long axis of the ellipsoid, are shown over time in figure \ref{FIG:FlowEllipsoid}~c) to h).
\new{
Here, the active stress triggers active deformation of the membrane which in turn triggers fluid flow inside the cell.
We note that the fluid velocity at the position of the membrane corresponds to membrane motion, which moves with local fluid velocity due to no-slip condition as described in section \ref{SEC:IBM}.
}
At the beginning the membrane contracts around the poles (left and right in figure \ref{FIG:FlowEllipsoid}~c)) due to the isotropic contractile active stress.
The contraction at the poles causes a rounding which triggers a flow field pointing away from the poles at the beginning.
Simultaneously, the membrane starts contracting around the equator, see figure  \ref{FIG:FlowEllipsoid}~c) and d).
In figure \ref{FIG:FlowEllipsoid}~d) four vortices are present around the equator and four at the corners of the figure.
After some time the contraction at the poles stops, see figure \ref{FIG:FlowEllipsoid}~e) and f) and only the contraction at the equator proceeds.
With progressive contraction a flow away from the midplane towards the poles develops.
As it is visible in figure \ref{FIG:FlowEllipsoid}~g) and h), the site of maximal velocity towards the poles is located at $x \approx \pm 10$.
Thus, it does not coincide with the center of the two spheroids pinching off at $x\approx\pm18$, but is rather shifted towards the equatorial plane at $x=0$.
Approaching the poles of the ellipsoid the velocity decreases.

\begin{figure}[!h]
	\centering
	\includegraphics[width=\textwidth]{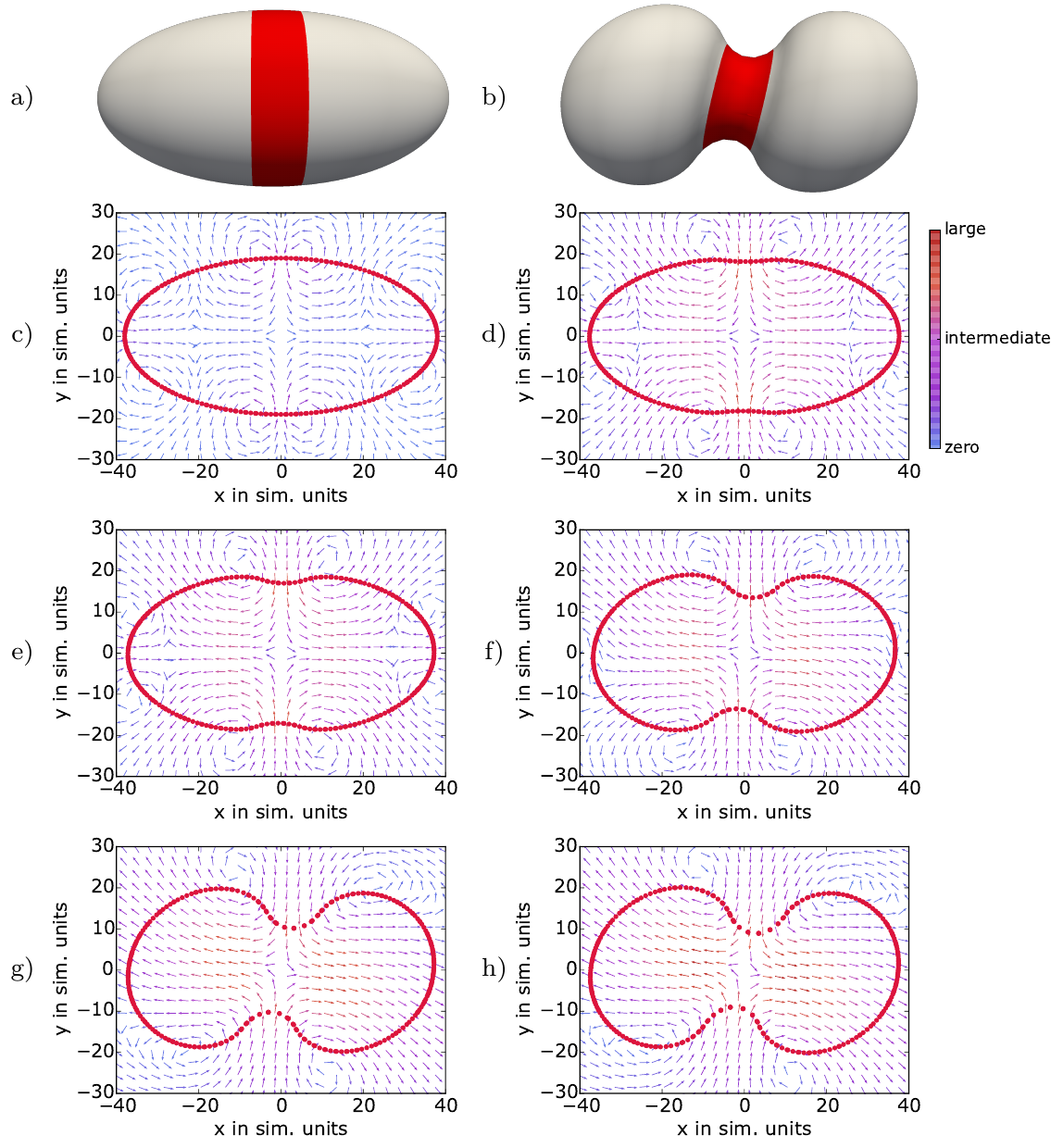}
	\caption{
			An elastic dividing cell membrane in shear flow.
			a) An increase in azimuthal active stress in the red shaded area of an ellipsoidal cell membrane triggers a local contraction as shown in b) while the externally driven shear flow also deforms the membrane.
			(c) \-- (h) show the outline of the deforming membrane (red nodes) in the central plane and the developing flow field (arrows) as perturbation to the shear flow over time.
			We here refer also to the Supplemental Video 1.
			The presence of the shear flow renders both the cell shape and the flow field asymmetric.
			\new{ Relative time is $t / t_{\text{max}}$ = a) 0, b) 1, c) 1.3$\times10^{-3}$, d) 0.07, e) 0.17, f) 0.52, g) 0.88, h) 1.0 .}
	}
	\label{FIG:FlowCellDivisionShear}
\end{figure}

\subsection{Dividing cell in shear flow}
\label{SEC:DividingShearFlow}

In the previous section we considered an initially quiescent fluid.
Here, we go one step further and include an externally driven flow interacting with the membrane.
We apply a shear flow with a shear rate $\dot{\gamma} \approx 1400$~s$^{-1}$.
All other parameters are the same as in the previous section.

The cell membrane as illustrated in figure \ref{FIG:FlowCellDivisionShear}~a) deforms now due to the local increase in active stress (red shaded area in figure a) and b)) but also due to the external shear flow.
This becomes visible by the non-symmetrically deformed membrane in figure \ref{FIG:FlowCellDivisionShear}~b).
Figures \ref{FIG:FlowCellDivisionShear}~c) to h) show the flow field relative to the shear flow, i.e., from each velocity vector the corresponding background flow is subtracted.
The time evolution of the cell shape and the flow field is also illustrated in the Supplemental Video 1.
In contrast to the previous section, the shear flow triggers an asymmetric deformation and in turn an asymmetric flow inside the cell.
The rounding together with the flow from the poles towards the equator is less pronounced (compare figure \ref{FIG:FlowCellDivisionShear}~d) to figure \ref{FIG:FlowEllipsoid}~d)).

The flow field inside a dividing cell suspended in a shear flow shows how the actively deforming membrane couples to a background flow and imposes perturbation on the shear flow.
Both the active stress present in the cell cortex as well as the external flow trigger membrane deformation.

\section{Conclusion}
\label{SEC:conclusion}

We presented a computational algorithm to compute the dynamical deformation of arbitrarily shaped active biological (cell) membranes 
embedded in a 3D fluid.
Active stresses in cells typically arise from the activity of motor proteins.
Constitutive equations for active stresses in membranes have been developed recently \cite{kruse_generic_2005,salbreux_mechanics_2017-1} in the framework of differential geometry and form the theoretical basis for our computational method.
The membranes are discretized by a set of nodes connected via flat triangles.
The key ingredient of our algorithm is the computation of the active force acting on each node starting from prescribed active stresses via a parabolic fitting procedure on the deformed membrane.
Besides active forces, the method also includes passive elastic forces derived from the well-established Skalak and Helfrich models for cell membranes.
In simple cases, the surrounding fluid can be considered a purely frictional medium, such that the membrane nodes follow simple overdamped dynamics in time.
For more realistic situations, we introduced a powerful and versatile coupling between the active membrane and the surrounding fluid via the Immersed-Boundary method.
This technique incorporates the full Navier-Stokes dynamics for the surrounding liquid, solved here via the Lattice-Boltzmann method.
Thus, our method allows us to go beyond the determination of equilibrium shapes of active elastic membranes and allows for simulations of dynamically deforming biological cells immersed in an external flow.

We successfully validated our algorithm for two distinct situations of an elastic, initially cylindrical membrane: (i) a local, Gaussian distributed active stress and (ii) a homogeneous active stress.
For (i) our numerical results are in excellent agreement with the analytically predicted deformation of \citet{berthoumieux_active_2014} and show convergence with increasing resolution.
Overdamped dynamics and IBM/LBM dynamics are in good agreement.
For (ii) we recovered both the buckling as well as the Rayleigh-Plateau-like instability predicted by \cite{berthoumieux_active_2014}.
Comparison to our own numerical solutions of the axisymmetric problem shows very good agreement with the full 3D algorithm also in the regime of large deformation not covered by \cite{berthoumieux_active_2014}.
In addition, our computations reveal the existence of a thus far unobserved non-axisymmetric instability in the case of extensile axial and azimuthal stresses.
In order to illustrate the versatility of our method, we analyzed the flow field inside an elastic, dividing cell membrane in shear flow.
This represents the first investigation of the dynamic two-way coupling between active deformations and externally driven fluid flow.
In this work, we considered temporally constant active stresses, but the inclusion of time-dependent active stresses computed, e.g., by a convection-diffusion model of active substances, is straightforwardly possible.

Our computational method significantly extends the range of physical problems to which existing active membranes theories can be applied.
First, it is not restricted to simple shapes such as cylinders or spheres (or small deformations thereof) allowing efficient and accurate treatment of arbitrary membrane shapes and deformations.
Second, our method couples the active membrane dynamics to the full Navier-Stokes dynamics of the surrounding fluid.
\new{
In the described LBM/IBM scheme a viscosity contrast of the inner and outer fluid \-- as it is well known in case of red blood cells \cite{freund_numerical_2014} \-- can furthermore be incorporated, which allows for a even more realistic model of living cells.
}
This opens up a wide range of applications in external flows such as active cells in the blood stream or active cellular compartments in cytoplasmic streaming flows which currently remain largely unexplored.
A particularly interesting application could be the formation of platelets from megakaryocytes which, according to a set of recent experiments \cite{bender_microtubule_2015,blin_microfluidic_2016}, crucially depends on the interplay between active processes and external flows.

\section{Acknowledgments}
C.~B.~thanks the Studienstiftung des deutschen Volkes for financial support and acknowledges support by the study program "Biological Physics" of the Elite Network of Bavaria.
This project is part of the collaborative research centre TRR225 (subproject B07) funded by the German Research Foundation (DFG).
We gratefully acknowledge computing time provided by the SuperMUC system of the Leibniz Rechenzentrum, Garching, as well as by the Bavarian Polymer Institute and financial support from the Volkswagen Foundation.

\clearpage

\renewcommand{\thesection}{A\arabic{section}}
\renewcommand{\theequation}{A\arabic{equation}}
\renewcommand{\thetable}{A\arabic{table}}
\renewcommand{\thefigure}{A\arabic{figure}}
\setcounter{figure}{0} 
\setcounter{section}{0}

\numberwithin{equation}{section}
\setcounter{equation}{0}

\vspace{0.75cm}
{\Large  Appendix}
\section{Membrane in thin shell formulation}
\label{app:thinShell}

In this appendix, we summarize the necessary basics and conventions of differential geometry on thin shells used in this work.
For a more detailed description we refer the reader to refs. \cite{green_theoretical_1954,deserno_fluid_2015}.
The 2D manifold in general is parametrized by two coordinates $s^1, s^2$.
We denote vector components on the manifold by Greek letters $\alpha, \beta = 1, 2$ and vector components in Euclidean space by Latin letters $i,j,k = x, y, z$. 
Moreover, we use the Einstein summation convention, i.e., double occurrence of an index in sub and superscript implies a sum over this index.
A partial derivative with respect to $s^{\beta}$ is denoted by a comma, i.e., for an arbitrary vector $\vect{v}$ this implies $v_{\alpha,\beta} = \partial v_\alpha / \partial s^{\beta}$.

The membrane in the undeformed state is parametrized by the vector
\begin{equation}
 \vect{X}(s^1,s^2,t).
\end{equation} 
From the local in-plane coordinates
\begin{equation}
   \vect{e}_{\alpha}= \vect{X}_{,\alpha}
   \label{EQ:inPlaneCoord}
\end{equation}
the metric tensor is defined by
\begin{equation}
  g_{\alpha\beta} = \vect{e}_{\alpha} \cdot \vect{e}_{\beta},
\end{equation}
with $g = \det(g_{\alpha\beta})$.
The inverse metric $g^{\alpha\beta}$ can be obtained by the expression
\begin{equation}
	g_{\alpha\gamma}g^{\gamma\beta} = \delta_{\alpha}^\beta
\end{equation}
with $\delta_{\alpha}^\beta$ denoting the Kronecker delta being 1 for $\alpha=\beta$ and 0 otherwise.
For a general vector $v_\alpha$, $v^\alpha$ or tensor $t_{\alpha\beta}$, $t_{\alpha}^{~\beta}$, $t^{\alpha}_{~\beta}$, $t^{\alpha\beta}$ subscript indices denote co-variant components and superscript indices denote contra-variant components.
An index can be raised by
\begin{equation}
	v^\alpha = v_\beta g^{\beta\alpha}~,~~~ t_{\alpha}^{~\beta} = t_{\alpha\gamma} g^{\gamma\beta}
\end{equation}
or lowered by
\begin{equation}
v_\alpha = v^\beta g_{\beta\alpha} ~,~~~ t_{\alpha\beta} = t_{\alpha}^{~\gamma} g_{\gamma\beta}.
\end{equation}
A surface element is defined by $dS = \sqrt{g}ds^1 ds^2$. 
The Christoffel symbols are given by
\begin{equation}
	\Gamma_{\alpha\beta}^{\gamma} = \frac{1}{2} g^{\gamma \delta} \left[ g_{\alpha\delta, \beta} + g_{\beta\delta, \alpha} - g_{\alpha\beta, \delta} \right].
\end{equation}
The in-plane coordinate vectors $\vect{e}_{\alpha}$ provide a local coordinate system together with the unit normal vector on the membrane
\begin{equation}
  \vect{n} = \frac{\vect{e}_1 \times \vect{e}_2}{\| \vect{e}_1 \times \vect{e}_2 \|}.
\end{equation}
Considering the local unit normal vector allows for the definition of the curvature tensor 
\begin{equation}
  C_{\alpha\beta} = -\vect{e}_{\beta, \alpha} \cdot \vect{n}.
\end{equation}
A co-variant derivative of an arbitrary vector $v^{\alpha}$ or tensor $t^{\alpha\beta}$ defined on the membrane is given by
\begin{align}
	\nabla_{\alpha} v^{\beta}  &= v^{\beta}_{~,\alpha} + \Gamma_{\alpha\gamma}^{\beta}v^{\gamma} \\
	\nabla_{\alpha} t^{\beta\gamma}  &= t^{\beta\gamma}_{~,\alpha} + \Gamma_{\alpha\delta}^{\beta} t^{\delta\gamma} + \Gamma_{\alpha\delta}^{\gamma} t^{\beta\delta}.
\end{align} 
On the membrane the Levi-Civita tensor is given by
\begin{equation}
	\epsilon_{\alpha\beta} = \sqrt{g}\begin{pmatrix} 0 & 1 \\ -1 & 0 \end{pmatrix}, ~~\epsilon^{\alpha\beta} = \frac{1}{\sqrt{g}}\begin{pmatrix} 0 & 1 \\ -1 & 0 \end{pmatrix}.
\end{equation}
An arbitrary vector $\vect{v}^\alpha$ defined on the manifold with respect to $\vect{e}_\alpha$ with $\alpha=1,2$ can be decomposed into an in-plane and a normal contribution
\begin{equation}
	\vect{v}^{\alpha} = t^{\alpha\beta} \vect{e}_{\beta} + t_n^{\alpha} \vect{n}
\end{equation}
with $t^{\alpha\beta}$ being the component of a tensor and $t_n^\alpha$ the component of a vector.

A connection from the in-plane coordinates to the Euclidean coordinates can be drawn by the expression
\begin{equation}
	\vect{e}_{\alpha} = e_{\alpha}^i \vect{E}_i
\end{equation}
with $e_{\alpha}^i$ being the $i$-th component of $\vect{e}_{\alpha}$ and $\vect{E}_i$ being the $i$-th Euclidean unit vector.
A three dimensional tensor $t_{ij}$ can be projected onto the membrane via
\begin{equation}
	t_{\alpha\beta} = t_{ij} e_{\alpha}^i e_{\beta}^j.
\end{equation}

External or internal forces may lead to a deformation of the membrane characterized by the deformation field $\vect{u}$.
The membrane in the deformed state is parametrized by
\begin{equation}
 \vect{X}^{\prime}(s^1,s^2) = \vect{X}(s^1,s^2) + \vect{u}(s^1,s^2).
 \label{EQ:deformedMembrane}
\end{equation}
We denote all vectors or tensors that are evaluated on the deformed surface by a prime.
Corresponding to the change of local coordinate vectors $\vect{e}_{\alpha}^\prime = \vect{X}_{,\alpha}^\prime$ and normal vector $\vect{n}^\prime$ both the metric tensor and the curvature tensor changes to
\begin{align}
g_\ab^\prime &= \vect{e}_\alpha^\prime \cdot \vect{e}_\beta^\prime \label{EQ:metricDeformed}\\
C_{\alpha\beta}^\prime &= -\vect{e}_{\beta, \alpha}^\prime \cdot \vect{n}^\prime.
\end{align}
The Christoffel symbols have to be computed using $g_\ab^\prime$ and the co-variant derivative becomes
\begin{align}
\nabla_{\alpha}^\prime v^{\beta} &= \partial_\alpha^\prime v^{\beta} + \Gamma_{\alpha\gamma}^{\prime\beta} v^{\gamma} \\
\nabla_{\alpha}^\prime t^{\beta\gamma}  &= \partial_{\alpha}^\prime t^{\beta\gamma} + \Gamma_{\alpha\delta}^{\prime\beta} t^{\delta\gamma} + \Gamma_{\alpha\delta}^{\prime\gamma} t^{\beta\delta}.
\label{EQ:covariantDeriv}
\end{align}

\section{Elastic in-plane surface stresses}
\label{SEC:Moduli}

In the following we compare the elastic in-plane surface stresses used in ref. \cite{berthoumieux_active_2014} to those obtained for Skalak energy density in equation \ref{EQ:Skalak}.
We consider the displacement vector in equation \ref{EQ:deformedMembrane} decomposed into axial and normal deformation
\begin{equation}
  \vect{u} = u_z \vect{e}_z + u_r \vect{n}.
  \label{EQ:deformationZN}
\end{equation}

For the given energy density in equation \ref{EQ:Skalak} the in-plane surface stresses are obtained by \cite{green_large_1960,lac_spherical_2004,daddi-moussa-ider_hydrodynamic_2017}
\begin{equation}
	t^{\alpha\beta}_{\text{SK}} = \frac{2}{J} \frac{\partial w^{\text{SK}}}{\partial I_1} g^{\alpha\beta} + 2 J \frac{\partial w^{\text{SK}}}{\partial I_2} g^{\prime\alpha\beta}
\end{equation}
with the invariants
\begin{align}
 I_1 &= g^{\alpha\beta} g^{\prime}_{\alpha\beta} - 2 \label{EQ:invariant1} \\
 I_2 &= \det(g^{\alpha\beta})\det(g^\prime_{\alpha\beta}) - 1, \label{EQ:invariant2}
\end{align}
and with $J = \sqrt{I_2 + 1}$.
Using equations \ref{EQ:deformedMembrane}, \ref{EQ:metricDeformed} and the deformation in equation \ref{EQ:deformationZN} we obtain for the metric on the deformed membrane in the limit of small deformations
\begin{equation}
	g^\prime_{\alpha\beta} = \begin{pmatrix}
		1 + 2 \partial_z u_z & 0 \\
		0 & 1 + 2 \frac{u_r}{R}
	\end{pmatrix}
\end{equation}
and for the in-plane surface stresses
\begin{align}
	t^{zz}_{\text{SK}} &= \frac{2}{3} \kappa_S \left[ (1+C)\partial_z u_z + C \frac{u_r}{R} \right] \\
	t^{\phi\phi}_{\text{SK}} &= \frac{2}{3} \kappa_S \left[ \frac{1}{R^2}C \partial_z u_z + (1+C) \frac{u_r}{R^3} \right].
\end{align}
These equations can be compared to the tensions in equation (11) of ref. \cite{berthoumieux_active_2014} for the elastic model used by \citet{berthoumieux_active_2014}.
The latter is based on the Hooke's law in 3D which is projected onto the membrane.
By comparing the stresses we find agreement in the limit of small deformations for the relation of the stretching modulus $S$ to the shear modulus used in equation \eqref{EQ:Skalak} of
\begin{equation}
	S = \frac{2}{3} \kappa_S
	\label{EQ:modulusSkappaS}
\end{equation}
and $C=1$.
This relation is used for the Green's function to match both elastic models.

We furthermore can compare the bending modulus used in \citet{berthoumieux_active_2014} to the one appearing in the Helfrich energy in equation \eqref{EQ:Helfrich} \cite{guckenberger_theory_2017}.
\citet{berthoumieux_active_2014} defines $B=Eh^3/(24 (1-\nu^2))$ with the Youngs modulus $E$ and the thin shell height $h$.
Comparing this to the expression $\kappa_B=Eh^3/(12 (1-\nu^2))$ of \citet{pozrikidis_effect_2001} we obtain the relation 
\begin{equation}
	B = \frac{1}{2} \kappa_B.
	\label{EQ:modulusBkappaB}
\end{equation}

\section{Ovedamped dynamics method and simulation analysis}
\label{SEC:method}

\subsection{Overdamped dynamics}

As a different approach than the LBM/IBM, we use a model program based on overdamped dynamics to solve for the final, equilibrium shape of the membrane in case of validation.
The resulting active and elastic forces $\vect{F}$ calculated for every node enter the equation of motion of the corresponding node $\vect{r}_c$ which is given by
\begin{equation}
	\vect{F} = \gamma \dot{\vect{r}}_c
\end{equation}
where $\gamma$ is a friction coefficient.
We solve the equations for all nodes using Euler integration scheme.
We fix the nodes at the boundaries of the cylinder by harmonic springs of strength $1000 \kappa_S$.
This results in nearly vanishing deformation at the boundary of the cylinder.

\subsection{Simulation analysis}

To obtain the shape shown for example in figure \ref{FIG:bellShaped}~a) or b) we average the final, radial deformation over all nodes at certain $z$-position.
Due to the averaging and inherent errors in the bending algorithm \cite{guckenberger_bending_2016} the deformation $u_r$ does not reach exactly zero far away from the perturbation in active stress, but show a constant offset of $5\times10^{-4}$, which we eliminate in the figures.

\section{Choice of reference neighbor does not influence simulation results}
\label{SEC:ReferenceNeighbors}

In section \ref{SEC:LocalCoordinate} we explain how the local coordinate system is built using the first neighboring node as reference for the first in-plane coordinate vector $\vect{e}_\xi$.
This fact is also important for the projection of the active stress into the local coordinate system, as mentioned in section \ref{SEC:specificationActive}.
In the following we show that the choice of the reference neighboring node is arbitrary and does not influence simulation results.

\begin{figure}[h!]
	\centering
	\includegraphics[width=\textwidth]{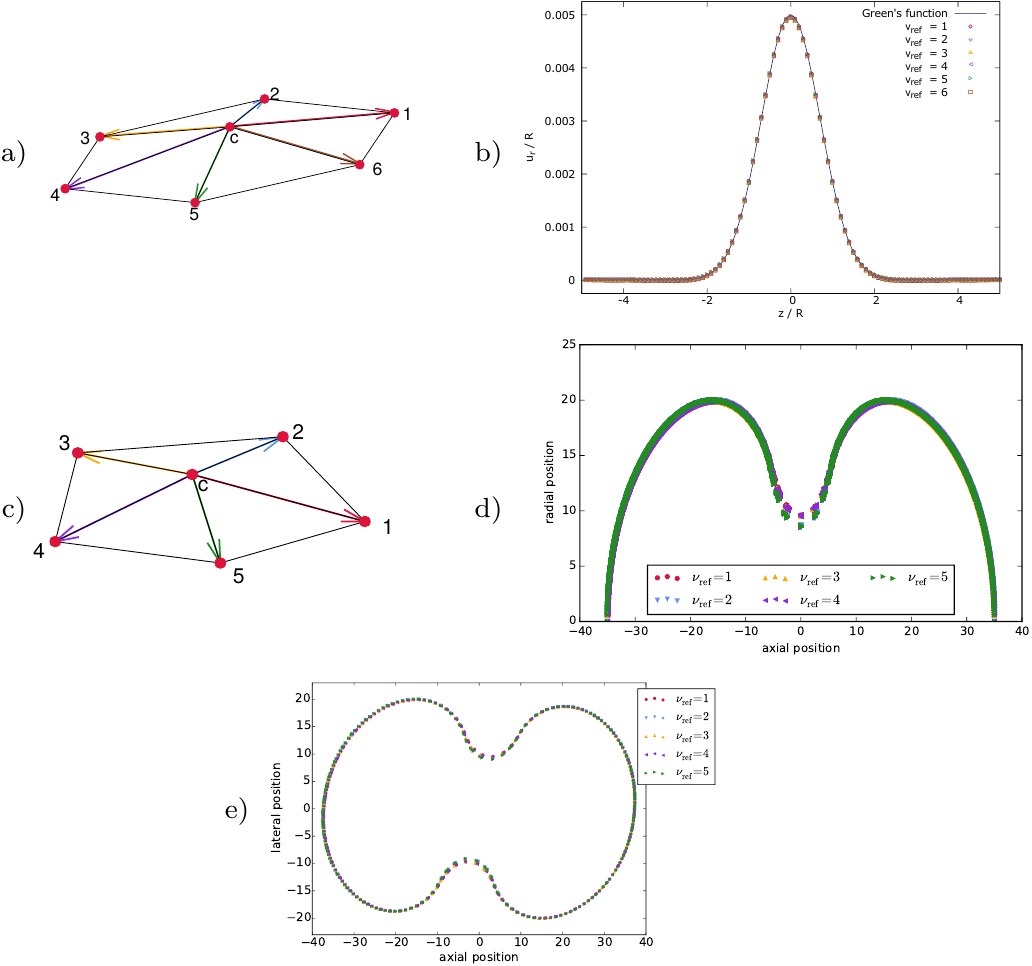}
	\caption{
	a) Color code for the neighbor node serving as reference to construct the local coordinate system on the cylindrical membrane at the site of the central node.
	b) 	Deformation obtained for a cylindrical membrane subjected to a Gaussian perturbation in active in-plane surface stress along $z$-direction for different reference neighbor nodes.
	The setup is identical to figure \ref{FIG:bellShaped}~c) with $T_\ia^z = 0.01$ and $\kappa_S = 1.0$.
	Obtained deformations are in very good agreement regardless of the choice of reference neighbor node.
	c) Neighbor node serving as reference to construct the local coordinate system on the ellipsoidal membrane at the site of the central node.
	d) Cell membrane shape for different reference neighboring node $\nu_{\text{ref}}$ at a given time corresponding to figure \ref{FIG:FlowEllipsoid}~g).
	Except for slight deviations in the region of largest curvature in case of neighboring node 2 and 5, all membrane shapes are in very good agreement.
	e) Membrane shapes in shear flow for different reference neighboring node $\nu_{\text{ref}}$ at a given time corresponding to figure \ref{FIG:FlowCellDivisionShear}~g) are in very good agreement.
	Thus, we prove evidence that the choice of the reference neighboring node does not affect simulation results.
	}
	\label{FIG:referenceNodes}
\end{figure}

Therefore, we first consider the cylindrical membrane subjected to a homogeneous perturbation in active in-plane surface stress along $z$-direction $T_\ia^z$.
We use exactly the setup analyzed in figure \ref{FIG:bellShaped}~c) with $T_\ia^z = 0.01$ and $\kappa_S = 1.0$.
We systematically change the reference neighbor node serving for construction of the local coordinate system as illustrated in figure \ref{FIG:referenceNodes}~a).
Figure \ref{FIG:referenceNodes}~b) shows that the deformation obtained in simulations is the same for every choice of reference neighbor node and agrees very well with the theoretically obtained Green's function.
In case of $T_\ia^z$ perturbation the in-plane derivatives of the active in-plane surface stress are  crucial and trigger the deformation.
Thus, figure \ref{FIG:referenceNodes} provides evidence for the correct calculation of derivatives regardless of the choice of reference neighbor node and furthermore shows the choice is arbitrary.

Furthermore, in order to take into account a more complex membrane geometry, we use the setup in figure \ref{FIG:FlowEllipsoid} and compare the dynamics for five different choices of reference neighboring node.
Choosing the sixth neighbor is not possible for the ellipsoidal geometry as the surface tiling requires at least 12 nodes with five neighbors only. 
We consider an initially ellipsoidal cell membrane endowed with active stress, which in azimuthal direction increases around the equator.
The increased active stress triggers the cell membrane to contract.
In figure \ref{FIG:FlowEllipsoid} we show the dynamically evolving flow field inside the cell.
Here, we redo the simulation and for each simulation we choose a different neighbor node to build the local coordinate system, which moves with the deforming membrane in time.
In figure \ref{FIG:referenceNodes}~c) we show the five different choices.

In figure \ref{FIG:referenceNodes}~d) we consider the point in time corresponding to figure \ref{FIG:FlowEllipsoid}~g) and show the radial position of all membrane nodes as function of the position along the axis.
All simulations with different reference neighbor show the same membrane shape and are in very good agreement.
We note that slight deviations occur at the site of strongest indentation, which results also in the strongest curvature.
At this position, the parabolic fit is not capable of covering the strongly deformed membrane shape completely and thus slight deviations occur.
In figure \ref{FIG:referenceNodes}~e) we do the same for a dividing cell in shear flow corresponding to figure \ref{FIG:FlowCellDivisionShear}~g).
We show the node positions within the plane containing the long axis of the cell and the shear axis of the external flow.
All simulations with different reference neighbor show the same membrane shape in shear flow and are in very good agreement.
Thus, figure \ref{FIG:referenceNodes}~d) and e) provide evidence that the choice of reference neighbor is indeed arbitrary and does not alter the simulation in case of a dynamically deforming membrane coupled to a suspending fluid.

\clearpage
\bibliographystyle{unsrtnat_c}
\bibliography{./bibliography.bib}

\newpage

\listoffigures

\newpage



\section*{Supplementary material}

# List of Figures



# Supporting Information - Computational modelling of active deformable membranes embedded in 3D flows

C. Bächer[1] and S. Gekle[1]

[1]Biofluid Simulation and Modeling, Theoretische Physik VI, Universität Bayreuth, Bayreuth, Germany

## S1  Numerics for an axisymmetric active elastic cylinder

Here we provide a detailed derivation of the numerical method used to obtain the axisymmetric shapes shown in figure 9 of the main text.

### S1.1  Axisymmetric, cylindrical shell

We consider a cylindrical shell under the assumption of axisymmetry. In the following we calculate the traction jump for the axisymmetric membrane for linear elasticity. The shell is parametrized using cylindrical coordiantes with $s^1 = z$ and $s^2 = \phi$ so that

$$\boldsymbol{X}(z,\phi) = R\cos\phi\,\boldsymbol{e}_x + R\sin\phi\,\boldsymbol{e}_y + z\boldsymbol{e}_z. \tag{S1}$$

We obtain for the in-plane coordinates $\boldsymbol{e}_z = (0,0,1)$, $\boldsymbol{e}_\phi = (-R\sin\phi, R\cos\phi, 0)$, and $\boldsymbol{n} = (-\cos\phi, -\sin\phi, 0)$. We follow Berthoumieux et al. [1] regarding conventions, thus the normal vector points inwards into the cylinder. The metric and curvature tensor become

$$g_{\alpha\beta} = \begin{pmatrix} 1 & 0 \\ 0 & R^2 \end{pmatrix} \quad C_{\alpha\beta} = \begin{pmatrix} 0 & 0 \\ 0 & R \end{pmatrix}. \tag{S2}$$

Due to axisymmetry the deformation can be written as

$$\boldsymbol{u} = u^z(z)\boldsymbol{e}_z + u^n(z)\boldsymbol{n}. \tag{S3}$$

and following equation (S3) we obtain the metric on the deformed surface using $g'_{\alpha\beta} = \boldsymbol{e}'_\alpha \cdot \boldsymbol{e}'_\beta$

$$g'_{\alpha\beta} = \begin{pmatrix} 1 + 2\partial_z u^z & 0 \\ 0 & R^2 - 2Ru^n \end{pmatrix}. \tag{S4}$$

Accordingly, the in-plane strain tensor can be calculated

$$u_{\alpha\beta} = \begin{pmatrix} \partial_z u^z & 0 \\ 0 & -Ru^n \end{pmatrix}. \tag{S5}$$

With the expressions above we can calculate the curvature tensor

$$C'_{\alpha\beta} = (\partial_\alpha \partial_\beta \boldsymbol{X}') \cdot \boldsymbol{n}' \tag{S6}$$

1

on the deformed membrane to be

$$C'_{\alpha\beta} = \begin{pmatrix} \partial_z^2 u^n & 0 \\ 0 & R - u^n \end{pmatrix}. \tag{S7}$$

The Levi-Civita tensor on the deformed surface for small deformations is

$$\epsilon'_{\alpha\beta} = (R + R\partial_z u^z - u^n) \begin{pmatrix} 0 & 1 \\ -1 & 0 \end{pmatrix}. \tag{S8}$$

The Christoffel symbols on the deformed membrane are

$$\Gamma^{z\prime}_{zz} \simeq \partial_z^2 u^z, \quad \Gamma^{z\prime}_{\phi\phi} \simeq R\partial_z u^n, \quad \Gamma^{\phi\prime}_{z\phi} \simeq -\frac{1}{R}\partial_z u^n, \quad \Gamma^{\phi\prime}_{\phi z} \simeq -\frac{1}{R}\partial_z u^n,$$

$$\Gamma^{\phi\prime}_{\phi\phi} = 0, \quad \Gamma^{\phi\prime}_{zz} = 0, \quad \Gamma^{z\prime}_{\phi z} = 0, \quad \Gamma^{z\prime}_{z\phi} = 0.$$

## S1.2 Force balance in axisymmetric formulation

In this section we consider the force balance equations in the case of an axisymmetric, cylindrical shell. The force balance (eqs. (11) - (12) in the main text) takes the following form in cylindrical coordinates

$$\nabla'_z t_a^{zz} + \nabla'_\phi t_a^{\phi z} - C'^z_z t_{na}^z - C'^z_\phi t_{na}^\phi + f_e^z = f^z \tag{S9}$$

$$\nabla'_z t_a^{z\phi} + \nabla'_\phi t_a^{\phi\phi} - C'^\phi_z t_{na}^z - C'^\phi_\phi t_{na}^\phi + f_e^\phi = f^\phi \tag{S10}$$

$$\nabla'_z t_{na}^z + \nabla'_\phi t_{na}^\phi + C'_{zz} t_a^{zz} + C'_{z\phi} t_a^{z\phi} + C'_{\phi z} t_a^{\phi z} + C'_{\phi\phi} t_a^{\phi\phi} + f_e^n = f^n. \tag{S11}$$

**Elastic forces** In the framework of our 3D simulations presented in the main text, elastic forces are computed by direct derivation of a discretized energy functional with respect to nodal positions. Here, we follow a different route and use elastic in-plane surface stresses in order to follow closely ref. [1]. Elastic properties of the shell are considered in the framework of Hooke's law for a three dimensional, elastic, isotropic solid with the stress tensor

$$\sigma_{ij} = \frac{E}{1+\nu}\left(e_{ij} + \frac{\nu}{1-2\nu}e_{kk}\delta_{ij}\right) \tag{S12}$$

depending in a linear fashion on the strain tensor $e_{ij} = u_{i,j} + u_{j,i}$ with $E$ being the Young's modulus and $\nu$ the Poisson ratio. From the bulk equation (S12) expressions for the intrinisc in-plane surface stress tensor and moments on the thin shell can be derived [1]

$$\bar{t}_{e\alpha\beta} = 2S\begin{pmatrix} \partial_z u^z - \nu\frac{u^n}{R} & 0 \\ 0 & \nu R^2 \partial_z u^z - u^n R \end{pmatrix} \tag{S13}$$

$$\bar{m}_{e\alpha\beta} = 2BR\begin{pmatrix} 0 & -R^2\partial_z^2 u^n - \nu u^n \\ \nu\partial_z^2 u^n + \frac{u^n}{R^2} & 0 \end{pmatrix}, \tag{S14}$$

with $S$ the stretching and $B$ the bending modulus. Following Salbreux and Jülicher [2] by using eq. (16) and (17) of [2] we calculate the in-plane surface stress and moment tensor via

$$t_e^{\alpha\beta} = \bar{t}_e^{\alpha\beta} + \frac{1}{2}\left(m^{\gamma\alpha}C_\gamma^{\ \beta} + m^{\gamma\beta}C_\gamma^{\ \alpha}\right) \tag{S15}$$

$$m_e^{\alpha\beta} = -\bar{m}_e^{\alpha\gamma}\epsilon_\gamma^{\ \beta} \tag{S16}$$

and obtain

$$t_e^{zz} = \bar{t}_e^{zz} \tag{S17}$$

$$t_e^{z\phi} = \bar{t}_e^{z\phi} \tag{S18}$$

$$t_e^{\phi z} = \bar{t}_e^{\phi z} \tag{S19}$$

$$t_e^{\phi\phi} = \bar{t}_e^{\phi\phi} - \frac{2B\nu}{R^3}\partial_z^2 u^n - \frac{2B}{R^5}u_n \tag{S20}$$

and
$$m_e^{zz} = -2B\partial_z^2 u^n + \frac{2B\nu}{R^2} u^n \tag{S21}$$
$$m_e^{z\phi} = 0 \tag{S22}$$
$$m_e^{\phi z} = 0 \tag{S23}$$
$$m_e^{\phi\phi} = -\frac{2B\nu}{R^2}\partial_z^2 u^n - \frac{2B}{R^4} u^n. \tag{S24}$$

We obtain for the normal surface stress
$$t_{ne}^z = \partial_z m^{zz} + 2(\partial_z^2 u^z) m^{zz} + (-\frac{1}{R}\partial_z u^n) m^{zz} + (R\partial_z u^n) m^{\phi\phi} \tag{S25}$$
$$t_{ne}^\phi = 0, \tag{S26}$$
and by inserting the in-plane moments we eventually get
$$t_{ne}^z = -2B\partial_z^3 u^n + \frac{2B\nu}{R^2}\partial_z u^n. \tag{S27}$$

Thus following
$$f_e^\beta = \nabla_\alpha' t_e^{\alpha\beta} + C_\alpha'^\beta t_{n,e}^\alpha \tag{S28}$$
$$f_e^n = \nabla_\alpha' t_{n,e}^\alpha - C_{\alpha\beta}' t_e^{\alpha\beta} \tag{S29}$$

we get for the elastic forces
$$f_e^z = 2S\partial_z^2 u^z - \frac{2S\nu}{R}\partial_z u^n \tag{S30}$$
$$f_e^\phi = 0 \tag{S31}$$
$$f_e^n = -2B\partial_z^4 u^n - \left(\frac{2B}{R^4} + \frac{2S}{R^2}\right) u^n + \frac{2S\nu}{R}\partial_z u^z \tag{S32}$$

**Active forces** We neglect the contributions of active moments and normal stress and use the active in-plane surface stress in the general form
$$t_{a\alpha}^\beta = \begin{pmatrix} t_a^z(z) & 0 \\ 0 & t_a^\phi(z) \end{pmatrix}. \tag{S33}$$

From the force balance in cylindrical coordinates we obtain the active force with components
$$f_a^z = \partial_z t_a^z - \frac{1}{R} t_a^z \partial_z u^n + \frac{1}{R} t_a^\phi \partial_z u^n \tag{S34}$$
$$f_a^\phi = 0 \tag{S35}$$
$$f_a^n = \partial_z^2 u^n t_a^z + \left(\frac{1}{R} + \frac{1}{R^2} u^n\right) t_a^\phi. \tag{S36}$$

Both elastic force and active force together determine the traction jump, which is required for simulations.

## S1.3 Numerical method

In order to determine the shape of the membrane in the steady state for the axisymmetric problem we perform an overdamped relaxation. We discretize the contour of the membrane (which is a line in case of the axisymmetric formulation) by a series of marker points. According to the traction jump in equations (S9) to (S11) we calculate the forces of the membrane using quintic splines for the derivatives of the deformation. Once the forces are determined, we consider the equations of motion in the overdamped limit for each membrane node
$$\boldsymbol{F} = \gamma\dot{\boldsymbol{r}}. \tag{S37}$$

This means we introduce a friction term in the equations of motion and neglect inertia. Equations of motion are solved by the Euler integration scheme. As boundary condition for the deformation we choose the first and second derivative to vanish on both ends.

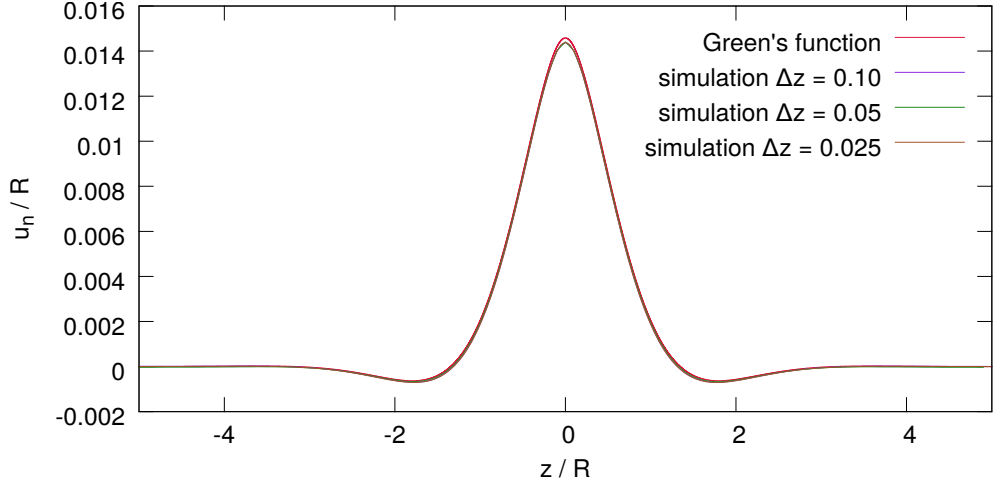

Fig. S1: Comparison of the numerical solution of the axisymmetric problem for three different resolutions with the Green's function analytically predicted by Berthoumieux et al. [1]. Numerics and analytics agree very well in the case of a small, local, active in-plane surface stress $T_a^\phi = -0.025$. Shapes are obtained for $R = 1$, $S = 1$, $b = 0.02$, and $g = 0$.

### S1.4 Results

In order to validate our axisymmetric numerical method, we first compare the numerical results to the predicted Green's function of Berthoumieux et al. [1]. For the introduction of the Green's function we refer to the section V.A of the main text. As in section V.B of the main text we apply an active in-plane surface stress tensor of the form

$$t_{a\alpha}^\beta = \begin{pmatrix} 0 & 0 \\ 0 & T_a^\phi \end{pmatrix} \delta(z). \tag{S38}$$

We compare axisymmetric simulations to the analytical Green's function in figure S1 for the parameter set $T_a^\phi = -0.025$, $T_a^z = 0$, $b = 0.02$, $g = 0$ for three different resolutions $\Delta z$. Figure S1 shows that our numerical method on the one hand does not depend on spatial discretization and on the other hand is in very good agreement with the theoretical expected deformation due to the Green's function given by ref. [1] including the shallow minima next to the main peak.

As in section V.D of the main text for the full 3D method, we next consider a finite, homogeneous, active in-plane surface stress $t_a \neq 0$ without any singular perturbation, i.e., $T_a^z, T_a^\phi = 0$. In this case Berthoumieux et al. [1] predicts two kinds of instabilities depending on the magnitude of active in-plane surface stress: for negative active in-plane surface stress beyond $t_a < -2\sqrt{\frac{3BS}{R^2}}$ a buckling instability takes place and for positive active in-plane surface stress they predict an instability for $t_a > 2S(1 - \nu^2)$. This allows us to perform simulations for different parameter sets $(g, b)$ and to construct a phase diagram classifying the final shape. The resulting phase diagram is shown in figure S2.

For negative, active in-plane surface stress $g < 0$ our axisymmetric numerics are in very good agreement with the predicted instability threshold for a broad range of relative bending moduli $b$. In the case of positive, active in-plane surface stress $g > 0$ the cylindrical shell contracts. Whereas the membrane in 3D shows an instability similar to a Rayleigh-Plateau instability of a liquid jet, in 2D the cylindrical membrane contracts homogeneously, since volume conservation is not enforced in the axisymmetric case. Beyond a certain threshold, the deformation becomes larger than the cylinder radius $R$. This serves us as a criterion for an instability. The Green's function found by Berthoumieux et al. [1] diverges for $g = 1.5$, which is beyond our instability threshold. The discrepancy may be explained by the fact that our criterion is more realistic than that of Berthoumieux et al. which is based on the Green's function only. Nevertheless, we note that our instability threshold does not depend on the bending modulus, as predicted by ref. [1].

In order to discuss the latter issue in further detail we show in figure S3 the final deformation $u_n^{\text{final}}/R$ depending on the relative, active in-plane surface stress $g$. We define the instability threshold by $|u_n^{\text{final}}| = R$. Beyond this threshold the deformation increases strongly, but our numerics do not cover the divergence predicted. However,

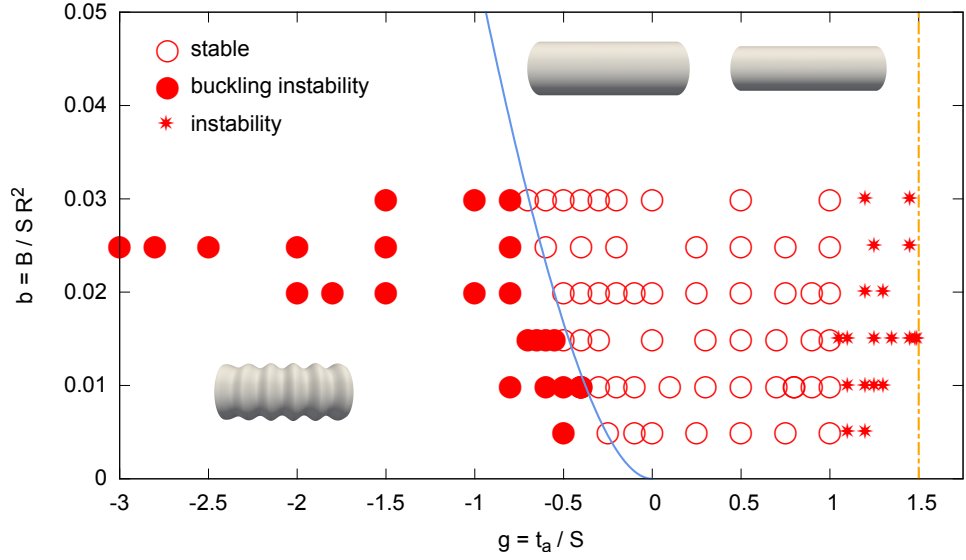

Fig. S2: Phase diagram of a cylindrical shell with homogeneous, relative, active in-plane surface stress $g = t_a/S$ and relative bending modulus $b = B/(SR^2)$ for axisymmetric simulations. The Poisson ratio is $\nu = \frac{1}{2}$. For negative active in-plane surface stress a buckling instability occurs. The simulations match the predicted [1] critical active in-plane surface stress very well. For positive active in-plane surface stresses an instability occurs characterized by a deformation larger than the cylinder radius which happens even before the threshold predicted by [1] (orange dotted line). Insets show membrane shapes corresponding to the different phases.

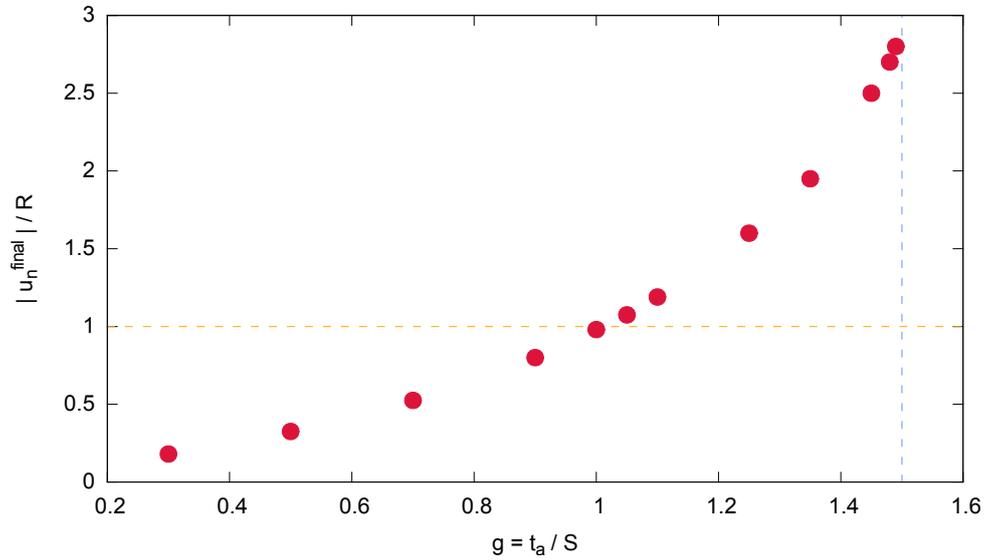

Fig. S3: Final deformation of a cylinder with homogeneous active in-plane surface stress in the axisymmetric simulations for a bending modulus $b \approx 0.015$. Near the threshold $g = 1.5$ the deformation increases strongly and nearly diverges (divergence not completely covered by numerics).

we have to note that the behavior beyond the threshold is not physical in a strict sense.

In conclusion, this section proves that our axisymmetric numerical simulations are in very good agreement with the predicted theory and consequently can be used for validation of the three dimensional method as done in figure 8 of the main text.



\end{document}